\documentstyle[11pt,epsfig]{article}
\begin{document}
\renewcommand{\thefootnote}{\fnsymbol{footnote}}
\sloppy
\newcommand{\lp}{\left(}
\newcommand{\rp}{\right)}
\newcommand \be  {\begin{equation}}
\newcommand \bea {\begin{eqnarray}}
\newcommand \ee  {\end{equation}}
\newcommand \eea {\end{eqnarray}}

\title{Punctuated vortex coalescence and
discrete scale invariance in two-dimensional turbulence}

\author{Anders Johansen$^1$, Didier Sornette$^{1,2,3}$, and Adam Espe
Hansen$^{4,5}$ \\
$^1$ Institute of Geophysics and Planetary Physics\\
University of California, Los Angeles, California 90095\\
$^2$ Department of Earth and Space Science\\
University of California, Los Angeles, California 90095\\
$^3$ Laboratoire de Physique de la Mati\`ere Condens\'ee\\
CNRS UMR6622 and Universit\'e des
Sciences, B.P. 70, Parc Valrose\\ 06108 Nice Cedex 2, France\\
$^4$ Niels Bohr Institute, \O rsted Laboratory, \\
Universitetsparken, 2100 Kbh. \O , Denmark\\
$^5$ Laboratoire de Physique Statistique, \'Ecole Normale Sup\'erieure \\
24 rue Lhomond, 75231 Paris, France}

\thispagestyle{empty}

\maketitle

\begin{abstract}

We present experimental evidence and theoretical arguments showing
that the time-evolution of freely decaying 2-d turbulence is
governed by a {\it discrete} time scale invariance rather than a continuous
time scale invariance. Physically, this reflects that
the time-evolution of the merging of vortices is not smooth but punctuated,
leading
to a prefered scale factor and as a consequence to log-periodic oscillations.
From a thorough analysis of freely decaying 2-d
turbulence experiments, we show that the number of vortices, their radius
and separation
display log-periodic oscillations as a function of time with an average
log-frequency of $\approx 4-5$ corresponding to a prefered scaling ratio of
$\approx 1.2-1.3$.
\end{abstract}

\newpage

\pagenumbering{arabic}

\section{Introduction}

Hydrodynamical turbulence is one of the major remaining challenges in the
physical sciences. It provides a fascinating phenomenology with both simple
universal scaling laws and complex system specific features.
Two-dimensional turbulence is of special interest, both for its applications in
astrophysics and geophysics and its theoretical properties. It is also
more amenable to detailed experimental, numerical and theoretical studies.
Its main characteristic property is the formation of and
interaction between coherent structures or vortices.

Kolmogorov first proposed his energy cascade model which was
later refined by many authors
to account for intermittency effects (see for instance \cite{Frisch}).
It remains to be assessed to what
extend this class of statistical phenomenological models is physically
correct and whether the the cascade picture can be taken
to represent real physical processes that occur genuinely in fluids.

2-d turbulence is distinct from 3-d turbulence by the conservation of vorticity
along fluid particle paths, when ignoring viscosity and forcing (there are no
vortex-stretching term in the 2-d equations).
An inverse energy cascade was conjectured by Kraichnan \cite{Kraichnan} and has
since been largely confirmed by simulations and experiments (see for
instance Ref.\cite{Frisch}
and references therein). This
inverse cascade does not appear to be characterised
by the presence of strong coherent vortices. Similarly,
if the direct cascade exists in 3-d turbulence,
vortex filaments are also rare events but, according to the so-called
hierarchical structure model \cite{She}, they nevertheless play an
important role.
These coherent structures are the most singular,
the most intermittent, and the rarest events in a turbulent medium;
these structures are also the most sensitive structures depending on the
environment. According to this theory, ``softer'' fluctuations are
organised statistically into a hierarchy which satisfies universal
statistical symmetry with scaling exponents controlled by the nature of these
most singular structures.

Returning to the 2-d inverse energy cascade, the coherent vortices are
clearly identified at the forcing scale, comparable to the system size.
Their statistical properties suggest that
the cascade is driven by a clustering mechanism
involving vortices of the same sign \cite{Paret}, in contrast to the
sequence of merging events producing larger and larger vortices occurring in
freely decaying 2-d turbulence which we analyse below.
In the inverse cascade, the energy is thus transported to larger scales by an
aggregation of vortices of the same sign while no apparent fusion is present,
in contrast to freely decaying 2-d turbulence. In addition, the inverse
cascade appears non-intermittent with regular Gaussian statistical properties
of the velocity increments.
The generalisation of our finding obtained
in a situation of vortex merging to forced 2-d turbulence where
vortices do not merge but aggregate and its relevance to the
energy cascade (inverse in 2-d and direct in 3-d) are thus conjectural and
remain
to be tested directly.

Here, we identify a novel signature of 2-d freely decaying turbulent flows
suggesting that vortex merging occur via a cascade process with a discrete
hierarchy of sizes. We document log-periodic oscillations, which are the
hallmarks of a discrete hierarchical structure with a prefered scaling ratio,
in the time evolution of the number of vortices, their radius and separation.

The theory of log-periodicity (and its associated complex exponents)
has advanced significantly \cite{revue} in
the last few years. Log-periodicity reflects a discrete scale invariance, i.e.,
the fact that dilational symmetry occurs only under magnification under
special factors, which are arbitrary powers $\lambda^n$ of a prefered
scaling ratio
$\lambda$. Log-periodicity has been studied in the eighties in relation
to various problems of physics embedded in discrete hierarchical systems.
In the context
of turbulence, shell models construct explicitly a discrete scale
invariant set of equations whose solutions are marred by unwanted
log-periodicities.
Only recently has it been realized that discrete scale invariance and its
associated complex exponents may appear ``spontaneously'' in Euclidean
systems, i.e., without the need for a pre-existing hierarchy (see
\cite{revue} for
references and discussion of Laplacian growth models,
rupture in heterogeneous systems, earthquakes, animals, financial crashes
among many other systems). In addition,
general field theoretical arguments \cite{SS}
indicate that complex exponents are to be expected generically for
out-of-equilibrium and quenched disordered systems.

The paper is organised as follows. In section 2, we discuss the plausibility
of log-periodicity in turbulence from the vantage of three theoretical
arguments
based on similarity analysis, mean field theory and the aggregation analogy.
In section 3 we describe
our statistical tools, in particular the Lomb periodogram technique and its
performance on synthetic tests with a detailed discussion of the role of
noise and the
problem of averaging. These considerations concerning
detection of log-periodic signals are offered to point-out
important pit-falls, which might explain why such signals have not previously
been reported. The experimental set-up, main experimental results
and our analysis of the experimental data are presented in section 4 while
section 5 concludes.

\section{Theory}

There is no established theory on 2-d freely decaying turbulence. All
available
approaches exhibit some weaknesses or incompatibilities with experiments.
Here we discuss three of the main theoretical approaches only to suggest
that log-periodicity is not unreasonable from a theoretical view point.

\subsection{Dimensional analysis and Similarity of the third kind}

Scaling laws are tools which can be used to quantify the complexity of
turbulent flows. Well-known examples include the Kolmogorov energy-cascade
from large scales to smaller scales in 3-d turbulence and from
small scales to larger scales in forced 2-d turbulence \cite{Frisch}, as
already mentioned.

For the decaying 2-d case, the flow carries coherent vortices
\cite{McW84}. Let us assume that the
number of vortices also obeys a scaling law. Using the $\Pi$ theorem of
dimensional analysis \cite{Barenblatt}, which states that the laws of physics
must be such as to be expressed using only dimensionless variables, it can
be written as
\be
n(t) \sim {1 \over E t^2} ~F(Re, {t \over t_L}, {t \over \tau},...)~.
\label{qlqml}
\ee
Here $E$ is the kinetic energy per area, $Re$ is the Reynolds number
inversely proportional
to the viscosity $\nu$, $t_L = L/\langle v \rangle$ is the typical time
scale for a vortex to
move over the system size $L$ and $\tau = \langle R \rangle/\langle v
\rangle$ is the typical
time for a vortex to move over a distance comparable to its typical radius.
That the first term of the r.h.s. is inversely proportional to $E t^2$ stems
from the fact that
$E$ is proportional to the inverse of a time-squared.

The simplest assumption \cite{Batchelor} is that the function $F$ goes to a
constant
when the viscosity $\nu \to 0$, leading to the prediction $n(t) \sim {1
\over t^2}$.
This corresponds to the so-called complete similarity of the first kind
\cite{Barenblatt} with respect to
the variables $Re$, $t/t_L$ and $t/\tau$.

The existence of a finite limit of $F$
was first questioned by L.D. Landau and A.M. Obukhov in the context of
fully developed turbulence in 3-d, on the basis of the
existence of intermittency -- large fluctuations of the energy dissipation rate
about its mean value.
Following the same argument and Barenblatt's classification leads to the
possibility
of an {\it incomplete similarity\/} in the variables $t/t_L$ and $t/\tau$. This
would require the absence of a finite
limit for $F(Re, {t \over t_L}, {t \over \tau},...)$ as $t/t_L \to 0$ or as
$t/\tau \to \infty$, and leads in the simplest case to the form
\be
n(t) \sim {1 \over E t^2} ~({t \over t_L})^{\alpha}~ ({t \over
\tau})^{\beta} \sim
{1 \over t^{\xi}}~,~~~~~~~{\rm with}~\xi \equiv 2-\alpha-\beta~,
\label{qlqqqml}
\ee
where $\alpha$ and $\beta$ are intermittency exponents.
If $\alpha$ and $\beta$ are real,
this corresponds to a similarity of the second kind \cite{Barenblatt}.

Incomplete self-similarity \cite{Goldenfeld,Dubrulle} may stem from a
possible dependence of the exponents on $Re$.
Dubrulle \cite{Dubthesis} (see also \cite{ET7}) has proposed to call the
case where $\alpha$ and/or $\beta$ is complex, leading to
\be
n(t) \sim {1 \over t^{\xi_R}} ~\cos [\xi_I \ln t + \psi]~,
\label{qlqqqqqml}
\ee
a similarity of the third kind, characterised by the absence of limit for
$F(Re, {t \over t_L}, {t \over \tau},...)$ and accelerated (log-periodic)
oscillations.

To our knowledge,  Novikov has been the first to point out in 1966 that
structure functions in 3-d turbulence should contain log-periodic oscillations
\cite{Novikov}. His argument was that if an unstable eddy in a turbulent
flow typically breaks up into two or three smaller eddies, but not into
$10$ or $20$ eddies, then one can suspect the existence of a prefered
scale factor, hence the log-periodic oscillations. This view does not seem
to be
born out by direct observation of 3-d flows where vortex filaments are very
rare
events and the cascade seems to occur via a different mechanism. The
dynamical occurrence of
a discrete hierarchy of velocity fluctuation remains to be investigated.
Notice however that our finding of a discrete hierarchy of merging events in
freely decaying 2-d turbulence is in a sense reverse to Novikov's picture.

Log-periodic oscillations have been repeatedly observed in structure factors
but do not seem to be stable and depend on the nature of the
global geometry of the flow and recirculation \cite{Frisch,Anselmet}
as well as the analysing procedure. As to be discussed in section \ref{tools},
the detection of log-periodic structures in data is a non-trivial task and is
rather sensitive to the implementation of the analysis.

\subsection{Mean field theory}

At times sufficiently large so that the individual mean time between two
collisions suffered by any one vortex is larger than the time needed
for a typical vortex to move the distance of its typical diameter,
we can use Trizac's mean field approach \cite{Trizac}
based on dimensional analysis. The fundamental equation resulting from this
analysis governs
the time evolution of the vortex density $n(t)$\,:
\be
{dn \over dt} = - \xi {n(t) \over t}~.
\label{ydfjxj}
\ee
It integration gives the usual power law solution
\be
n(t) \sim t^{-\xi}~,
\label{hskq}
\ee
where $\xi$ is empirically found equal to $0.72 \pm 0.03$ while theoretical
arguments
suggest $\xi=1$ \cite{Pomeau,Sire} or $\xi = 0.8$ \cite{Trizac}.

It is important to recognise that (\ref{ydfjxj}) represents an averaging
over highly intermittent and ``non-smooth'' processes\,: most of the time,
the dynamics of the freely decaying
2-d turbulence is dominated by the mutual advection of the vortices, while rare
dissipative
events of energetic vortex merging punctuate this dynamics. Thus, in
reality, the
time dynamics is not scale free as expression (\ref{ydfjxj}) would lead us
to believe.
There are rather well-defined instants at which the coalescence proceeds,
corresponding to a discrete time evolution for the number of vortices, and
the occurrence of theses instants have to obey a scale invariance in time
dilation in order to recover (\ref{ydfjxj}) on average.

Let us now go beyond this average description and attempt to describe this
punctuated dynamics by introducing time scales in the simplest way such as
to recover the continuous time scale invariance (CTSI) upon averaging\,: we
hypothesise that the sudden coalescence can be accounted for by breaking
partially the CTSI captured by expression (\ref{ydfjxj}) into a weaker {\it
discrete} time scale invariance (DTSI). Quantitatively, this amounts to
replacing (\ref{ydfjxj}) by expressing $dt$ as a discrete time step
approximation with should also be scale invariant\,:
\be
dt = \alpha t~,
\ee
with $\alpha$ a characteristic scale ratio. Then, (\ref{ydfjxj}) transforms
into
$$
{n(t+\alpha t) - n(t) \over \alpha t} = - \xi {n(t) \over t}~,
$$
leading to
\be
n(\lambda t) = (1-(\lambda -1) \xi) ~n(t)~,~~~~~~{\rm where}~~\lambda
\equiv 1 + \alpha~.
\label{gqjq}
\ee
Expression (\ref{gqjq}) is now a discrete scale invariant relation for
$n(t)$. Its general solutions are power laws $t^{-s_m}$ like (\ref{hskq}),
where the exponents $s_m$ now depend on $\alpha$ and can acquire an
imaginary part. It is this imaginary part which is the signature of a
discrete hierarchy of time scales reflected in log-periodic oscillations\,:
\be
s_m = - {\ln (1-(\lambda -1) \xi) \over \ln \lambda} + i m{2 \pi \over \ln
\lambda}~,
\label{fqjjd}
\ee
where $i=\sqrt{-1}$ and $m$ is an integer. The real solution $s_0$
recovers
the value $\xi$ in the limit $\alpha \to 0, \lambda \to 1$, where the
discrete scale
symmetry recovers the continuous scale symmetry.

We see that, for $\lambda > 1$, $s_0 > \xi$. As an example, take the
experimentally determined value $\xi = 0.72$, which gives $s_0 = 0.80$ for
$\lambda = 1.13$ and $s_0 = 1.01$ for $\lambda = 1.4$. For $\lambda = 1.2$,
we get $s_0 = 0.85$. This suggests an explanation for the difference
between prediction from theory and experimental and numerical data\,:
the measurements that assume CTSI underestimate the true value that is
impregnated by the existence the punctuated dynamics leading to DTSI.

In this formalism, one can take into account the presence of fluctuations
in the rate
of vortex mergers by replacing (\ref{gqjq}) by
\bea
n(\lambda_1 t_0) &=& (1-(\lambda_1 -1) \xi_1) ~n(t_0)\\
n(\lambda_2 \lambda_1 t_0) &=& (1-(\lambda_2 -1) \xi_2) ~n(\lambda_1 t_0)\\
...\\
n(\prod_{i=1}^N \lambda_i  t_0) &=& n(t_0) ~\prod_{i=1}^N (1-(\lambda_i -1)
\xi_i) ~.
\label{hfjls}
\eea
These equations describe the fact that the hierarchy of time scales at
which the vortex
mergers occur do not need to obey an exact hierarchy but can exhibit
fluctuations.
Looking for a power law behaviour $n(t) \sim t^{-s}$, we find that the {\it
typical}
value for $s$ is given by \cite{SS}
\be
s_m = - {\langle \ln (1-(\lambda -1) \xi) \rangle \over \langle \ln \lambda
\rangle}   +
i m{2 \pi \over \langle \ln \lambda \rangle} ~,
\label{fqqqjjd}
\ee
where the averaging $\langle . \rangle$ is performed over the
distributions of $\lambda$ and $\xi$.

\subsection{Linear theory of aggregation}

Benzi et al. \cite{Benzi} and McWilliams \cite{McWilliams} have noted the
analogy between the vortex merging in  turbulence and the kinetic theory
of colloidal aggregation \cite{Vandongen}. Starting from
a population $n_1(t=0)$ of monomers, the kinetic equations of aggregation read
\bea
{dn_1 \over dt} &=& - n_1 ~\sum_{i=1}^{\infty} J_{1,i} ~n_j\\
{dn_j \over dt} &=& \sum_{i=1}^{j-1} J_{i,i-j} ~n_i n_{j-i} - n_j
~\sum_{i=1}^{\infty} J_{i,j} ~n_j~.
\label{gqjqk}
\eea
The first equation expresses that
the population of monomers decreases by coalescence with all possible
species. The second
equation expresses the competition between the formation of a $j$-mer by
coalescence of
an $i$-mer and a $j-i$-mer and its destruction by coalescence with all
possible species.
The $J_{i,j}$ express the rates at which these coagulation occur. When
diffusion motion
dominates, we have
\be
J_{i,j} \sim (R_i + R_j) (D_i + D_j)~,
\ee
where $R_i$ (resp. $D_i$) is the radius (resp. diffusion coefficient) of
the $i$-mer.
From Stokes law and Einstein's relationship, $D \sim 1/R$ the kernel
are homogeneous\,:
$J_{\lambda i, \lambda j} = J_{i,j}$. In this case, the $J_{i,j}$ can be
taken constant
and (\ref{gqjqk}) reduces to
\bea
{dn_1 \over dt} &=& - A n_1 n\\
{dn \over dt} &=& {A \over 2} n_1^2 - {1 \over 2} n^2~,
\label{jkqkqmlqm}
\eea
where $n(t) = \sum_{i=1}^{\infty} n_j$ is the total population. It is then
straightforward
to check that $n_1(t) \sim n(t) \sim 1/t$ at large $t$.

A fundamental physical ingredient in this approach is the quadratic pair-wise
nature of the interactions described by the r.h.s. of the equations
(\ref{gqjqk},\ref{jkqkqmlqm}), which embodies the idea that coagulation is a
stochastic process requiring the chancy collision of {\it two}
particles that can come from arbitrary position in space. We argue quite
reasonably that this is not the correct situation for vortex fusion in
freely decaying 2-d turbulence due to the confinement that a 2-d space
represents.
Experiments (see for example figure 3 of
Ref.\cite{Hansen}) show that, between two merging events, vortices move
as in a flexible ``cage'' of neighbouring vortices. Thus, the relevant
rate of fusion is not proportional to the square of the number of
vortices but to the first power, since each vortex can only merge with
those vortices
at its perimeter that do not evolve significantly over long period of times.
In other words, there is not an efficient mixing and hence the mean-field
approximation underlying eq.s (\ref{gqjqk}) and (\ref{jkqkqmlqm}) breaks down.

As a consequence, we can write down a {\it linear} integro-differential
coupled equation for the time
evolution of the vortex populations, in close analogy to the scaling
theory of linear fragmentation of
Ref.\cite{Redner,Cheng}. A remarkable feature of this approach is that the
analysis can be carried out for arbitrary scaling kernels.

It is convenient to define the distribution
$\Phi(E,t)$ of vortex energies $E$.
For time independent vorticities $\omega$, the vortex radius $R$ is
obtained from its energy
from the relation $E \approx \omega^2 R^4$. The strategy is
to characterise the
vortex energy population $\Phi(E,t)$ by the moments $m_{\alpha} =
\int_0^{\infty} l^{\alpha} \phi(l) dl$, where we use
the scaling ansatz of the vortex
energy/size distribution
\be
\Phi(E,t) = {1 \over \epsilon^2} \phi({E \over \epsilon})~,
\label{hqlmqmlq}
\ee
and $\epsilon(t)$ is the characteristic (time-dependent) vortex energy.
The exponent $-2$ is required by (approximate) energy conservation.
Note that calculating a moment corresponds to taking
 the Mellin transform of $\phi(l)$. This implies
that if we know $m_{\alpha} $, we are able to retrieve $\phi(l)$ by taking
the inverse Mellin transform.

Adapting from Cheng and Redner \cite{Cheng}, we obtain
the equations of evolution for the
vortex size population.
Assuming that the fusion rate between two vortices of similar energy
$E$ is a power law $E^{\beta}$ and calling
$b(x)$ the probability that the ratio between the final vortex
energy and each of the initial fusing vortex be $x$, we substitute the scaling
ansatz (\ref{hqlmqmlq}) in the linear Smoluchowsky aggregation equations
and obtain
two separate equations with a separation constant $\omega$, one for the time
dependence of $\epsilon(t)$ and the other for the scaling function $\phi$.
In terms of the moments $m_{\alpha}$ of $\phi$, we get the following
recurrence equations\,:
\be
m_{\alpha + \beta} = \omega {1- \alpha \over L_{\alpha} - 1} m_{\alpha}~,
\label{yhqkkqlq}
\ee
and
\be
L_{\alpha} = \int_0^1 x^{\alpha} b(x) dx~,
\ee
From the term ${1 \over L_{\alpha} - 1}$ in (\ref{yhqkkqlq}), we see that if
there exist a value $\alpha^*$ such that
\be
L_{\alpha^*} = 1~,
\label{qlq}
\ee
 then all moments with
$\alpha > \alpha^*$ will become infinite. Provided reasonable analyticity
conditions hold, it follows that the value $\alpha^*$
is a pole of $m_{\alpha}$. Taking the inverse Mellin transform of
$m_{\alpha}$ then allows us to
get $\phi(l)$ and, using the existence of the pole at $\alpha^*$, this
immediately predict that
\be
\phi(l) \sim l^{-(1+ \beta + \alpha^*)}~,
\ee
 One can
check that for a large variety of $b(x)$, there are solutions of
(\ref{qlq}) with complex exponents $\alpha^*$. This signals the existence
of log-periodic corrections to a pure power law distribution of vortex
sizes. Such a scenario has been documented experimentally in the reverse
situation of fragmentation in Ref.\cite{Ouillon}.
This, in turn, leads to a complex separation constant $\omega$, and thus to a
complex dynamical exponent $\xi$ for the time evolution.

\section{Methodology of the data analysis \label{tools}}

\begin{figure}
\begin{center}
\epsfig{file=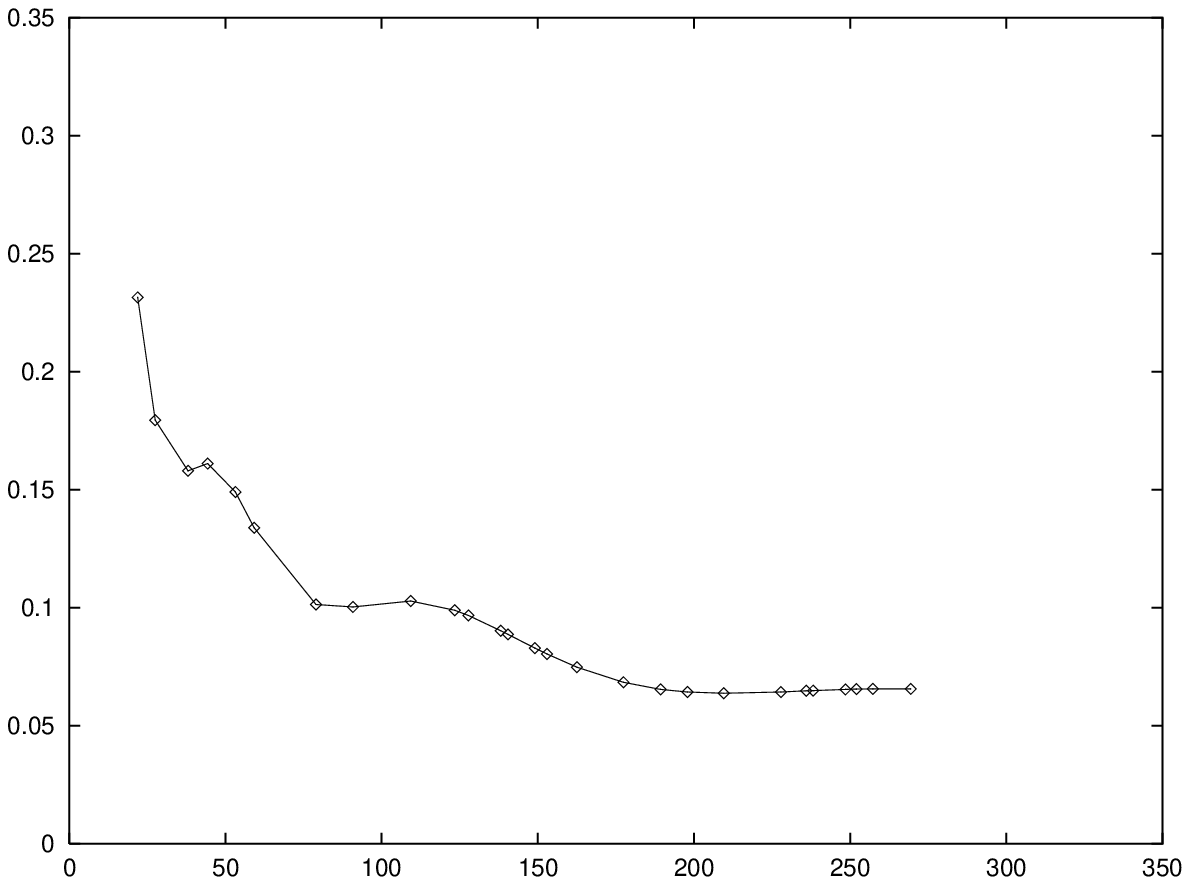,width=7.5cm}
\epsfig{file=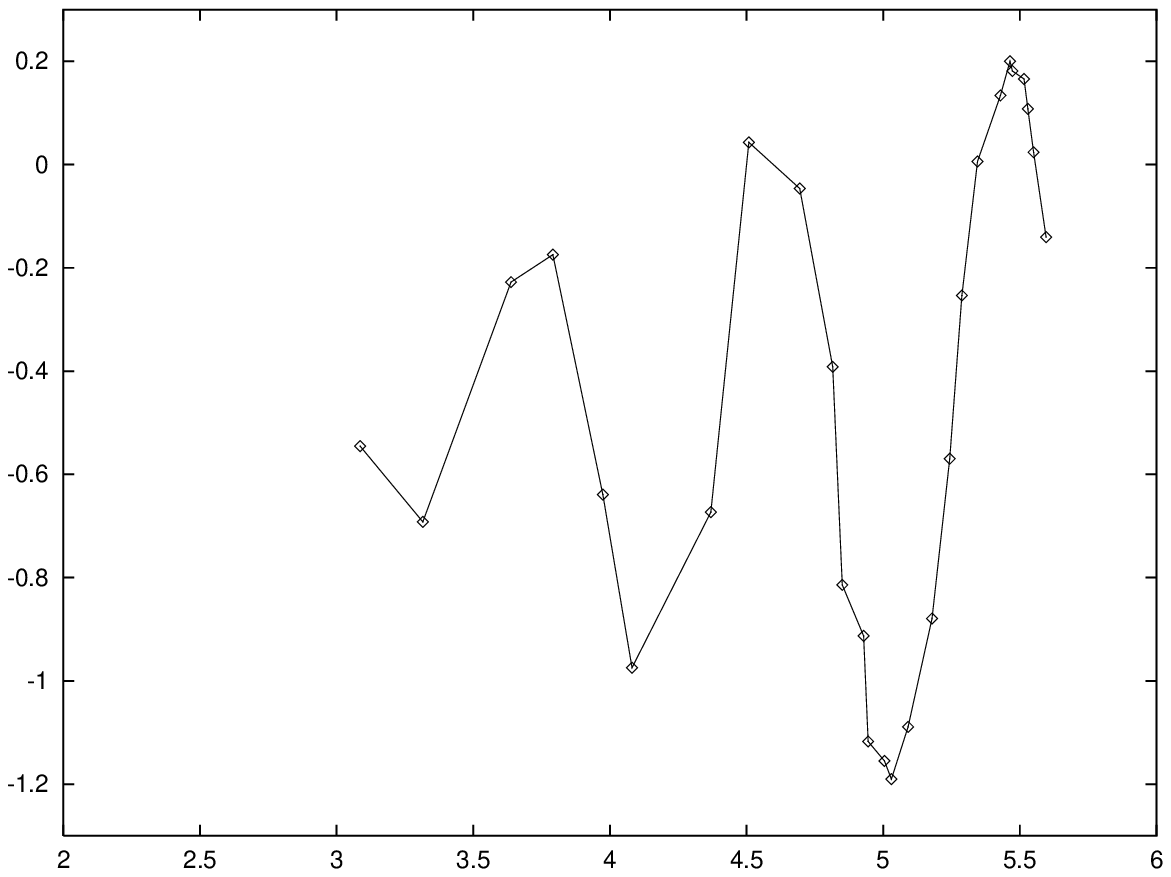,width=7.5cm}
\caption{\protect\label{art1} First synthetic data set and it's logarithmic
derivative.}

\vspace{1cm}

\epsfig{file=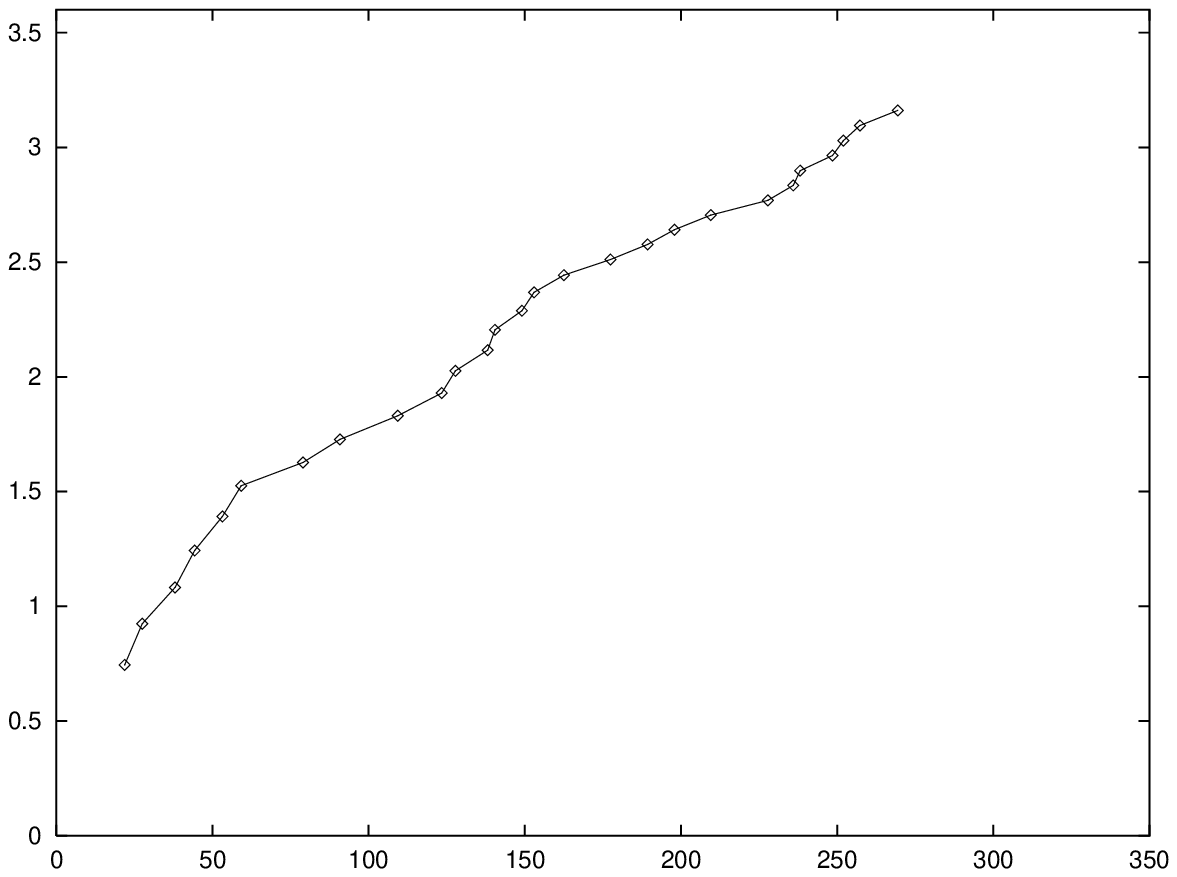,width=7.5cm}
\epsfig{file=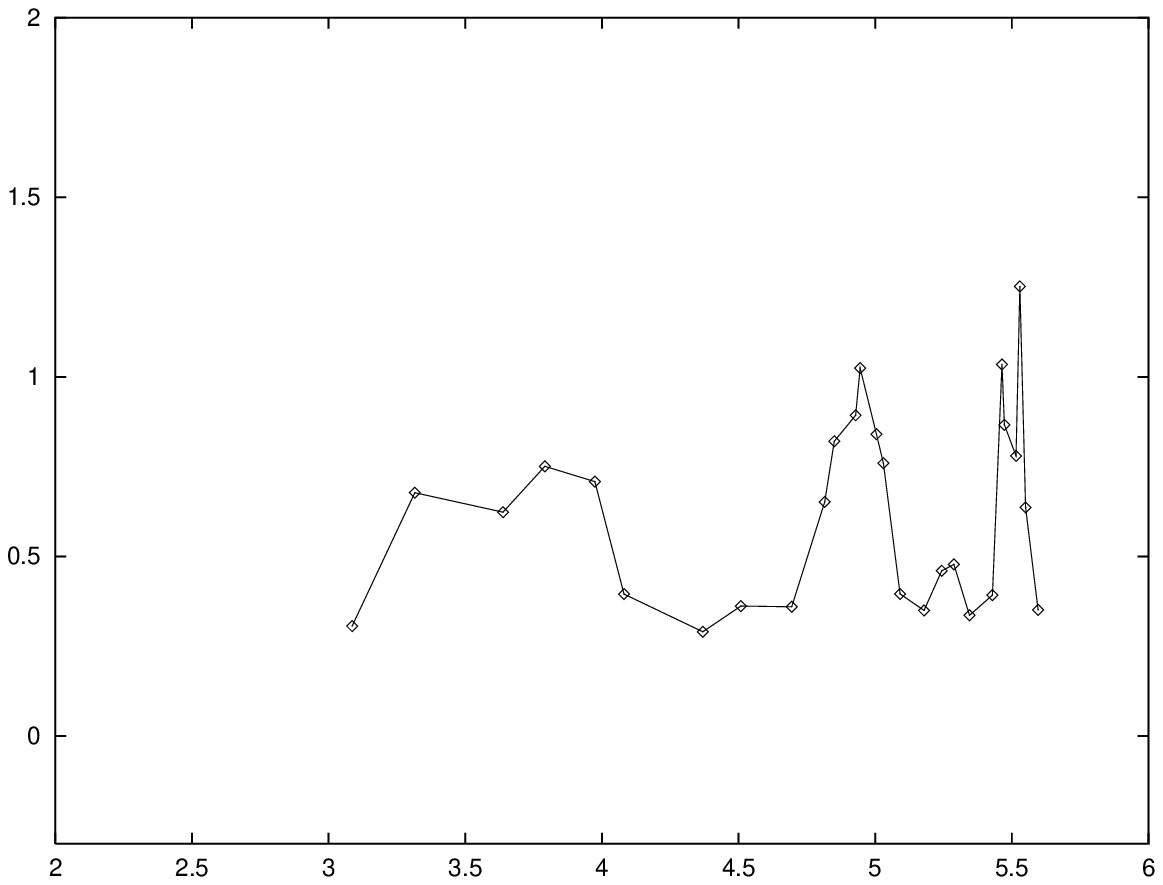,width=7.5cm}
\caption{\protect\label{cumart1} The cumulative distributions of the data set
in figure \protect\ref{art1} and it's logarithmic derivative.}

\vspace{1cm}

\epsfig{file=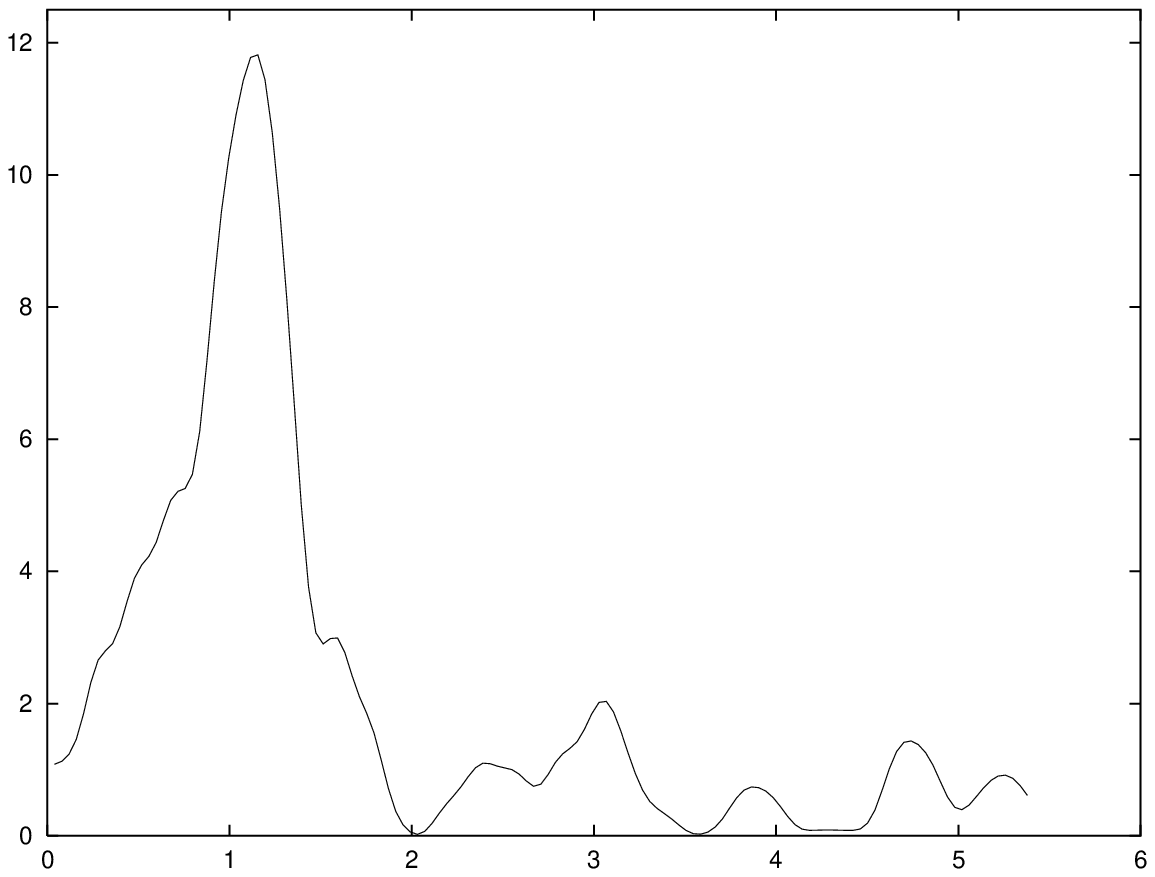,width=7.5cm}
\epsfig{file=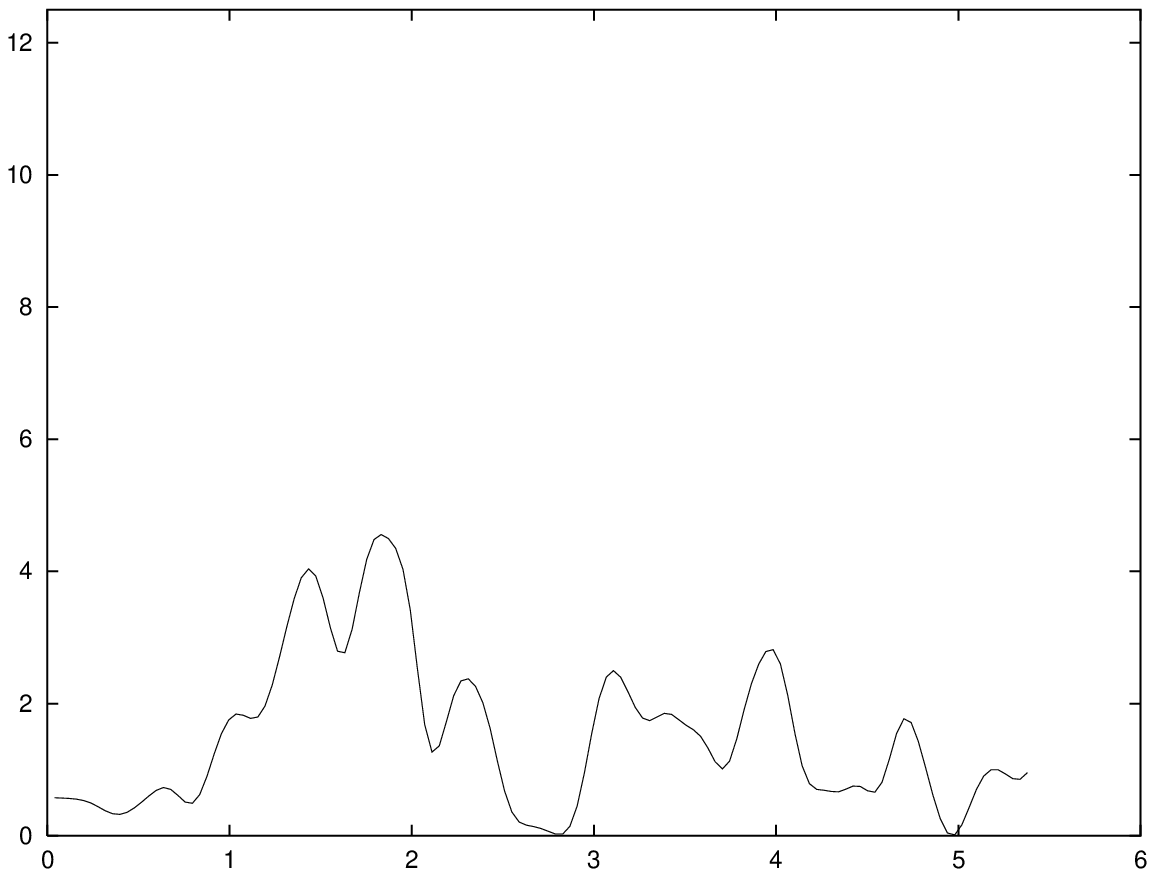,width=7.5cm}
\caption{\protect\label{lombart1} Lomb periodograms of the derivatives in
figures \protect\ref{art1} (left) and \protect\ref{cumart1} (right). }

\end{center}
\end{figure}

\begin{figure}
\begin{center}
\epsfig{file=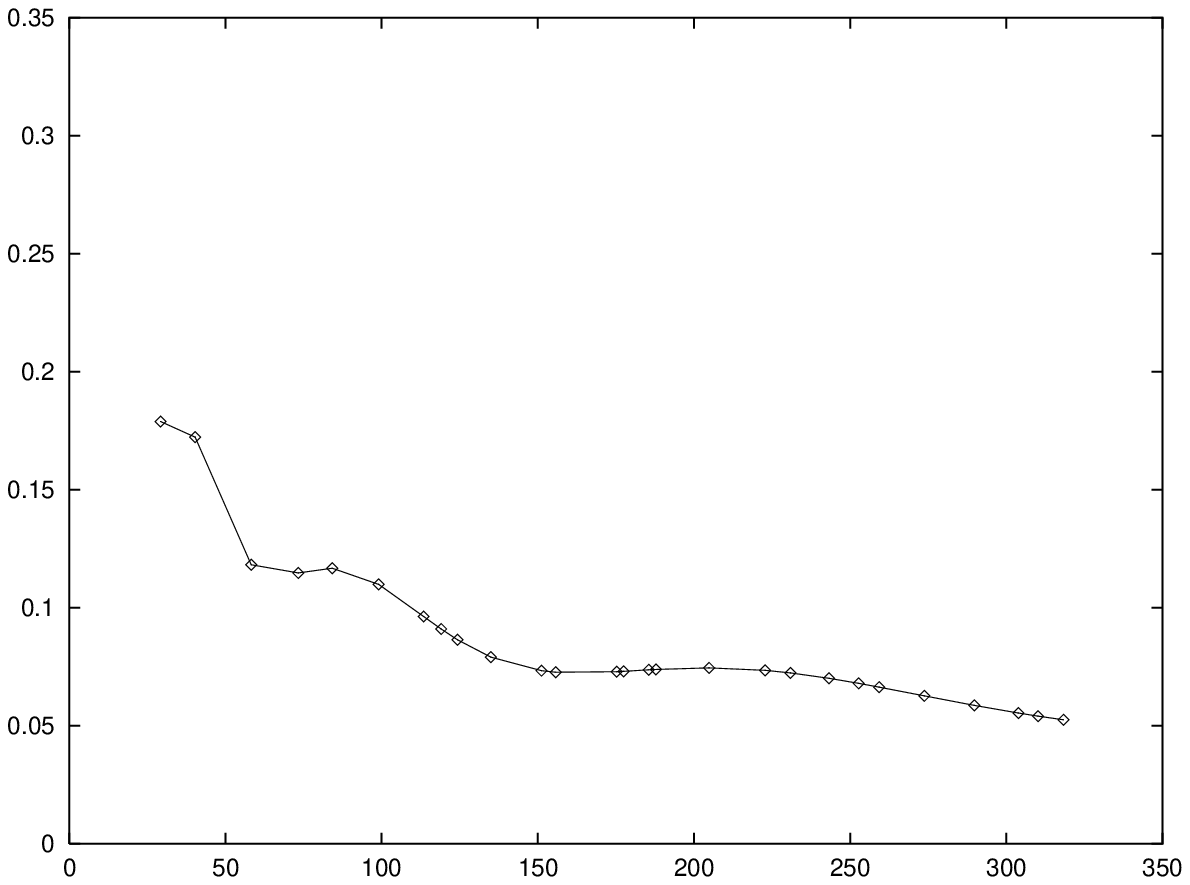,width=7.5cm}
\epsfig{file=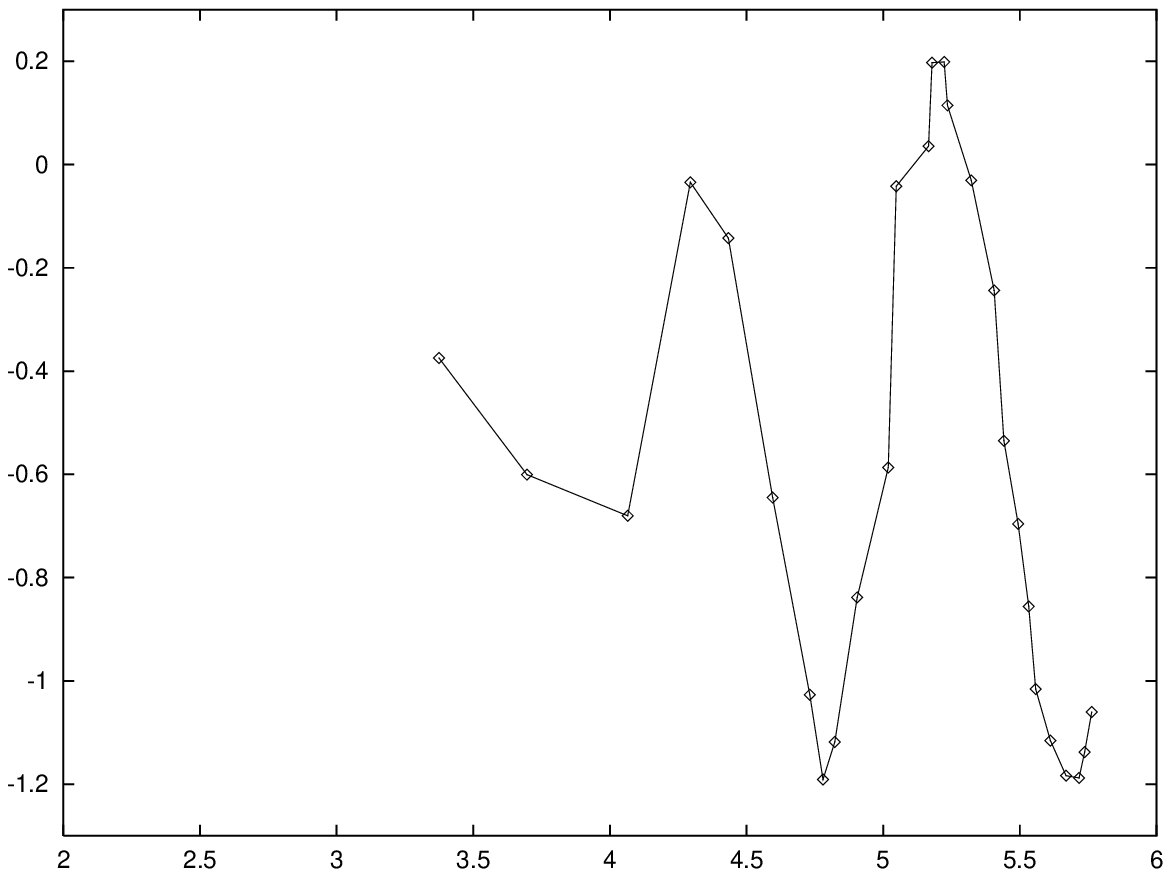,width=7.5cm}
\caption{\protect\label{art2} Second synthetic data set and it's logarithmic
derivative.}

\vspace{1cm}

\epsfig{file=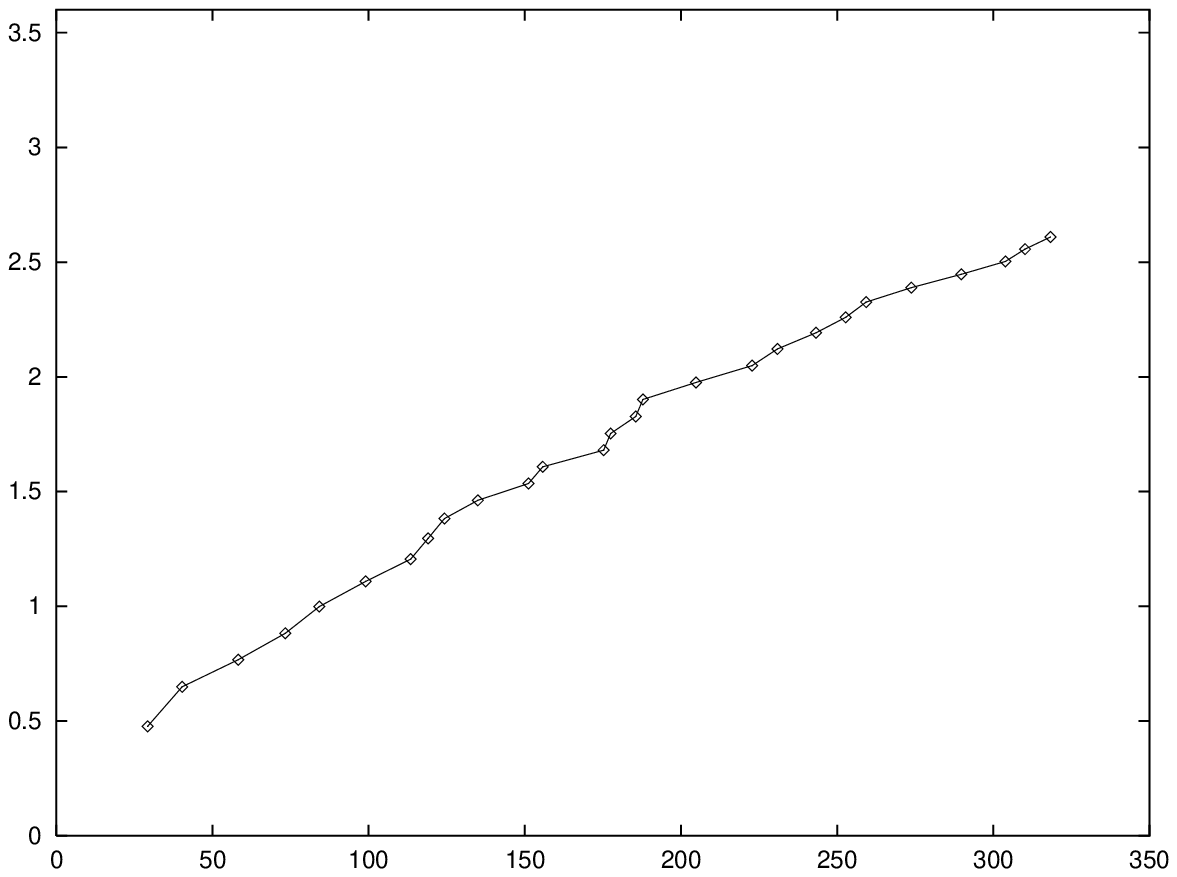,width=7.5cm}
\epsfig{file=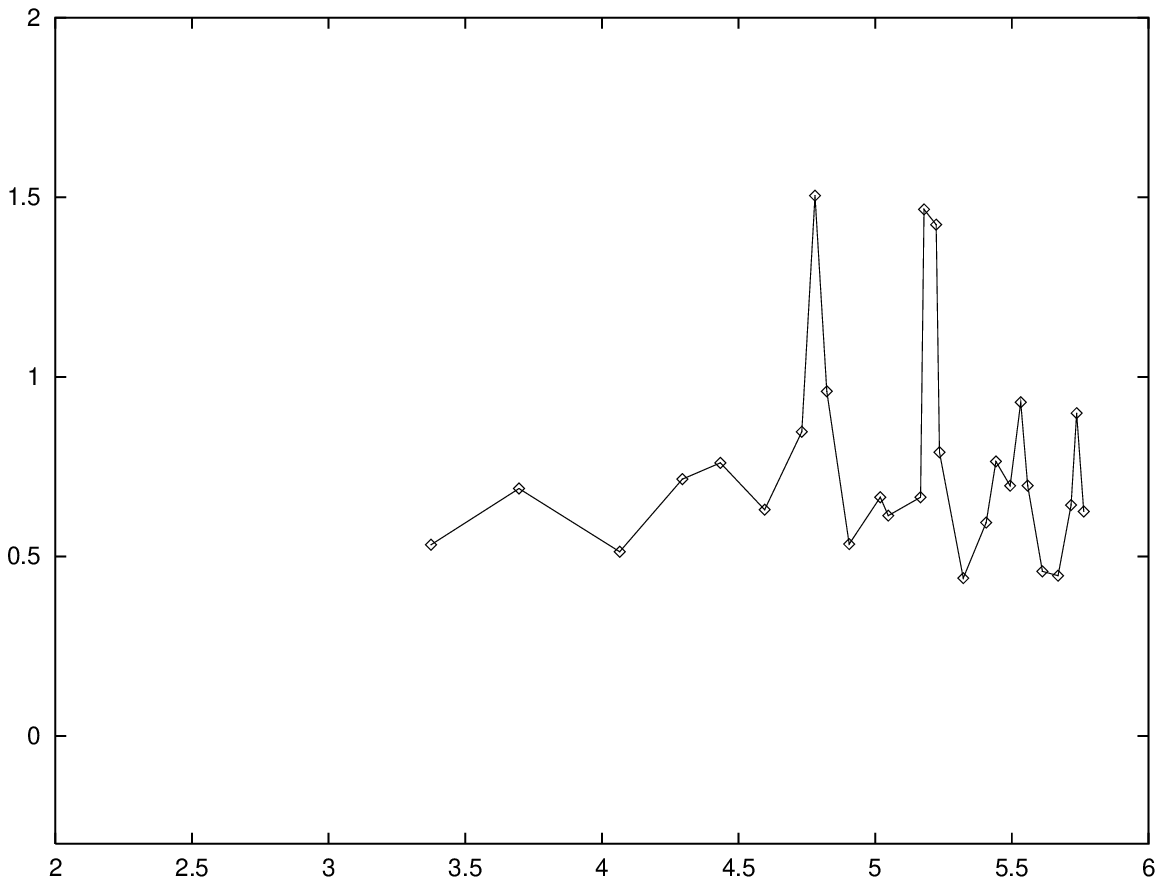,width=7.5cm}
\caption{\protect\label{cumart2} The cumulative distributions of the data set
in figure \protect\ref{art2} and it's logarithmic derivative.}

\vspace{1cm}

\epsfig{file=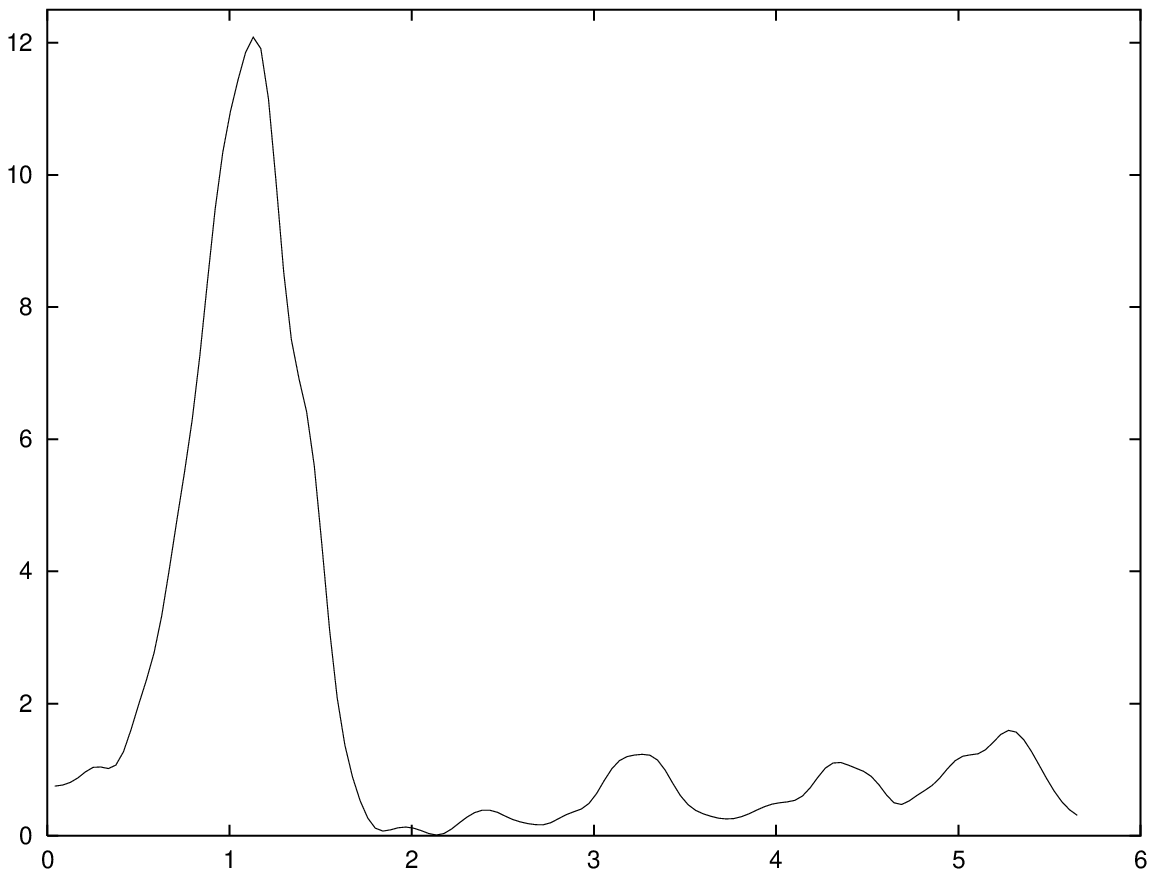,width=7.5cm}
\epsfig{file=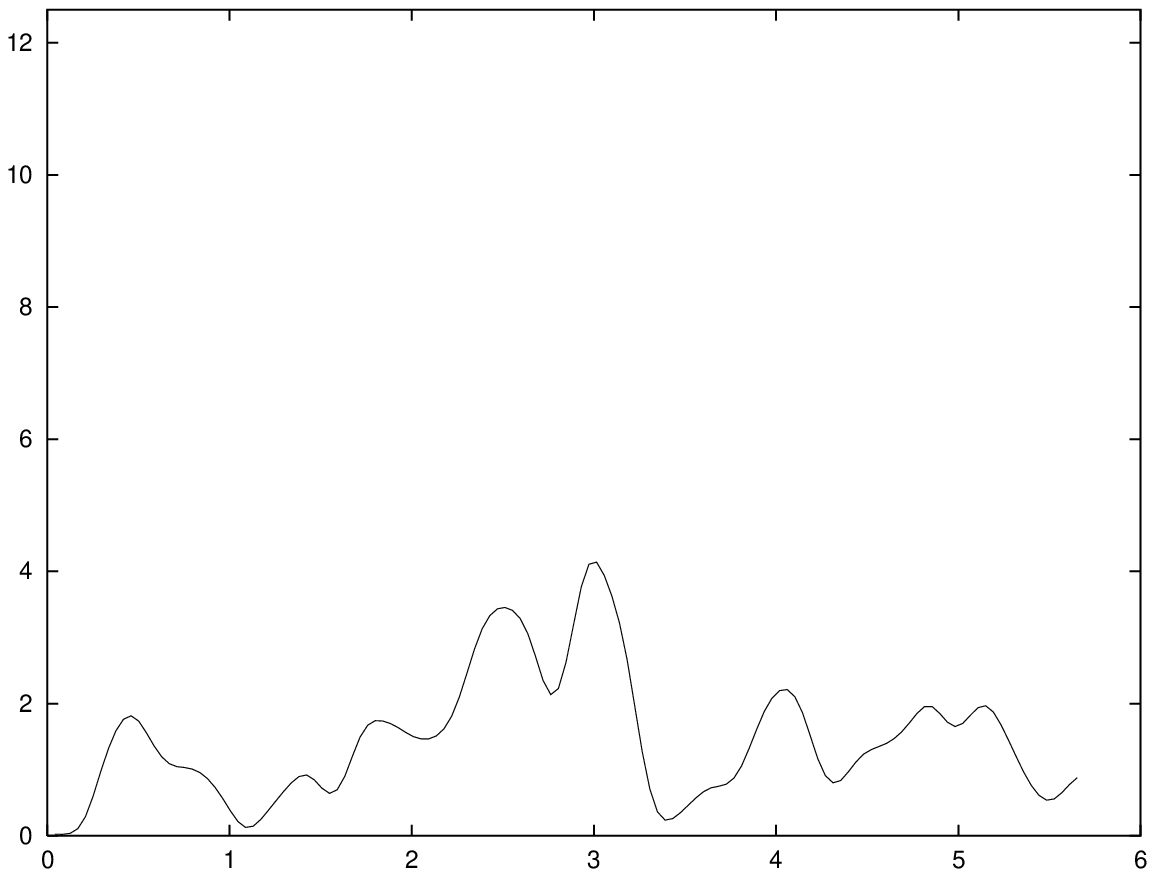,width=7.5cm}
\caption{\protect\label{lombart2} Lomb periodograms of the derivatives in
figures \protect\ref{art2} (left) and \protect\ref{cumart2} (right). }

\end{center}
\end{figure}

\begin{figure}
\begin{center}
\epsfig{file=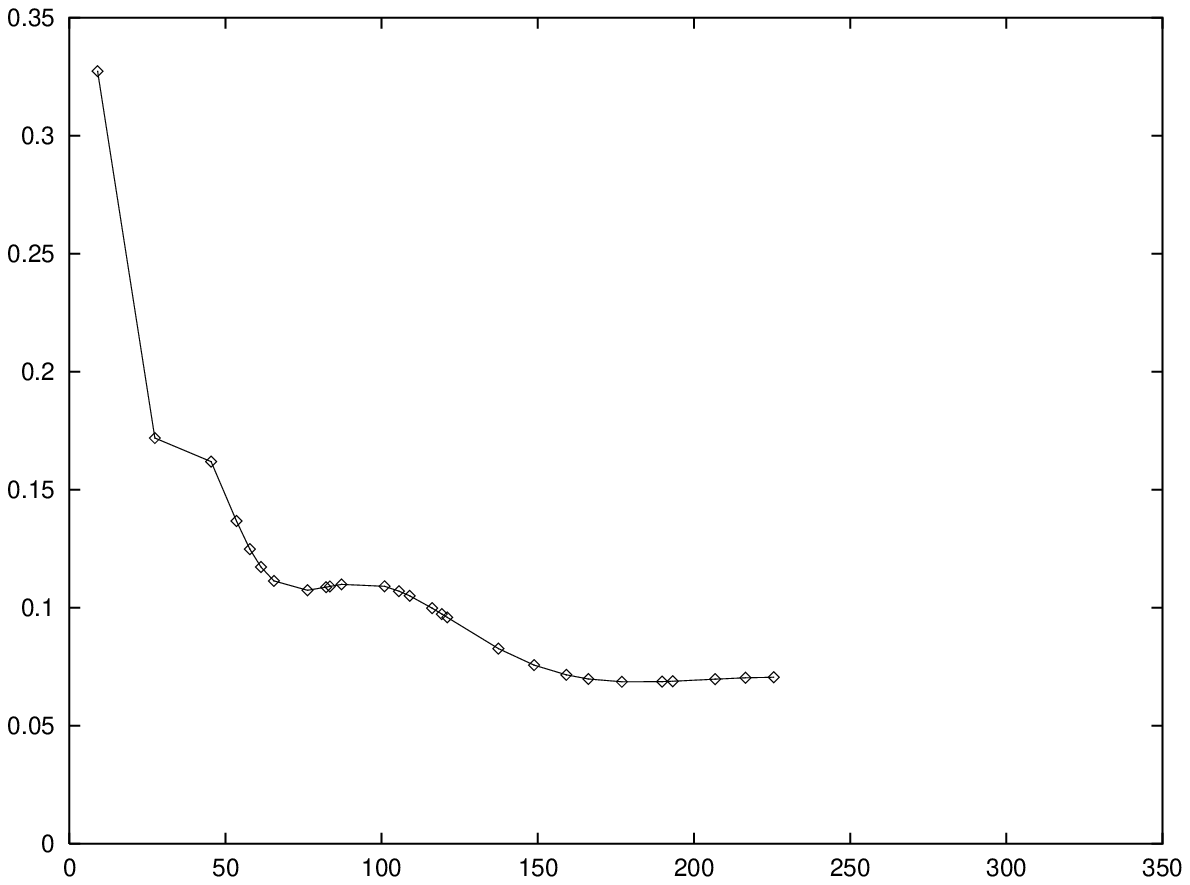,width=7.5cm}
\epsfig{file=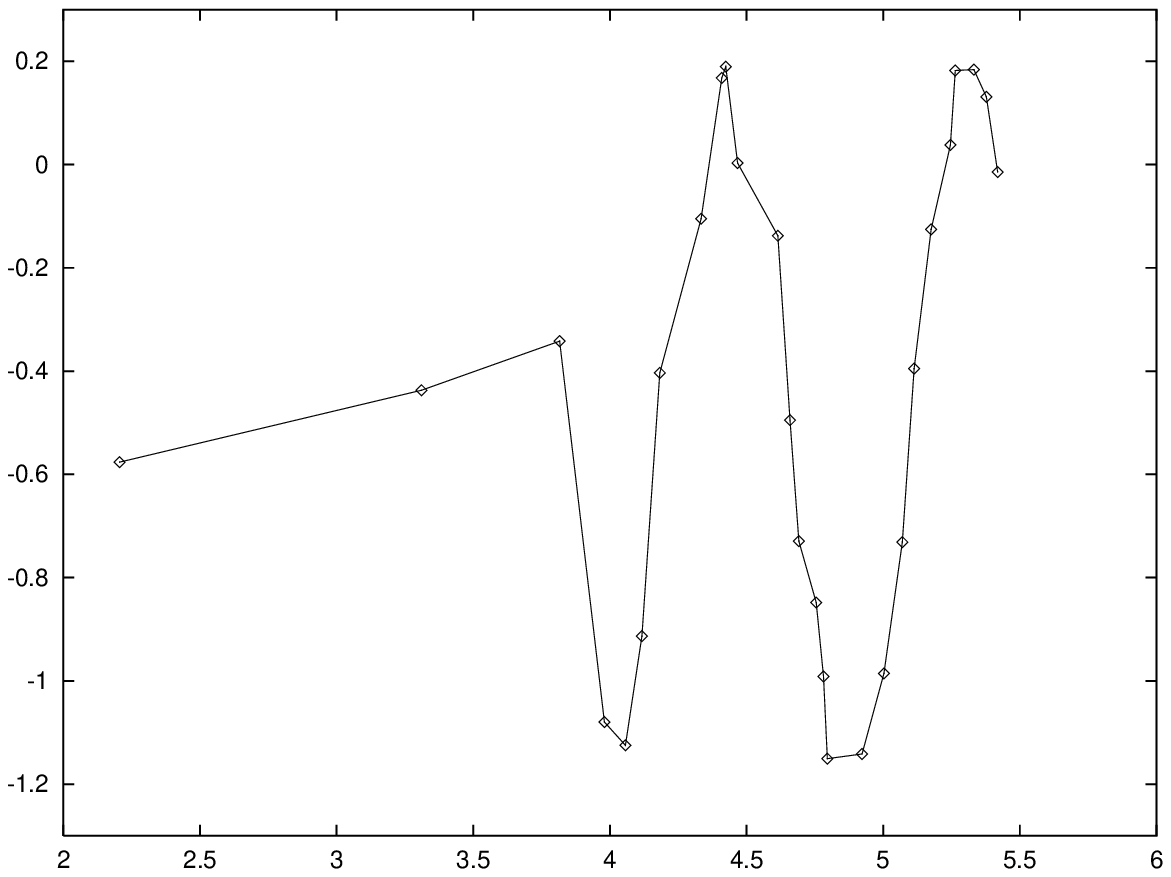,width=7.5cm}
\caption{\protect\label{art3} Third synthetic data set and it's logarithmic
derivative.}

\vspace{1cm}

\epsfig{file=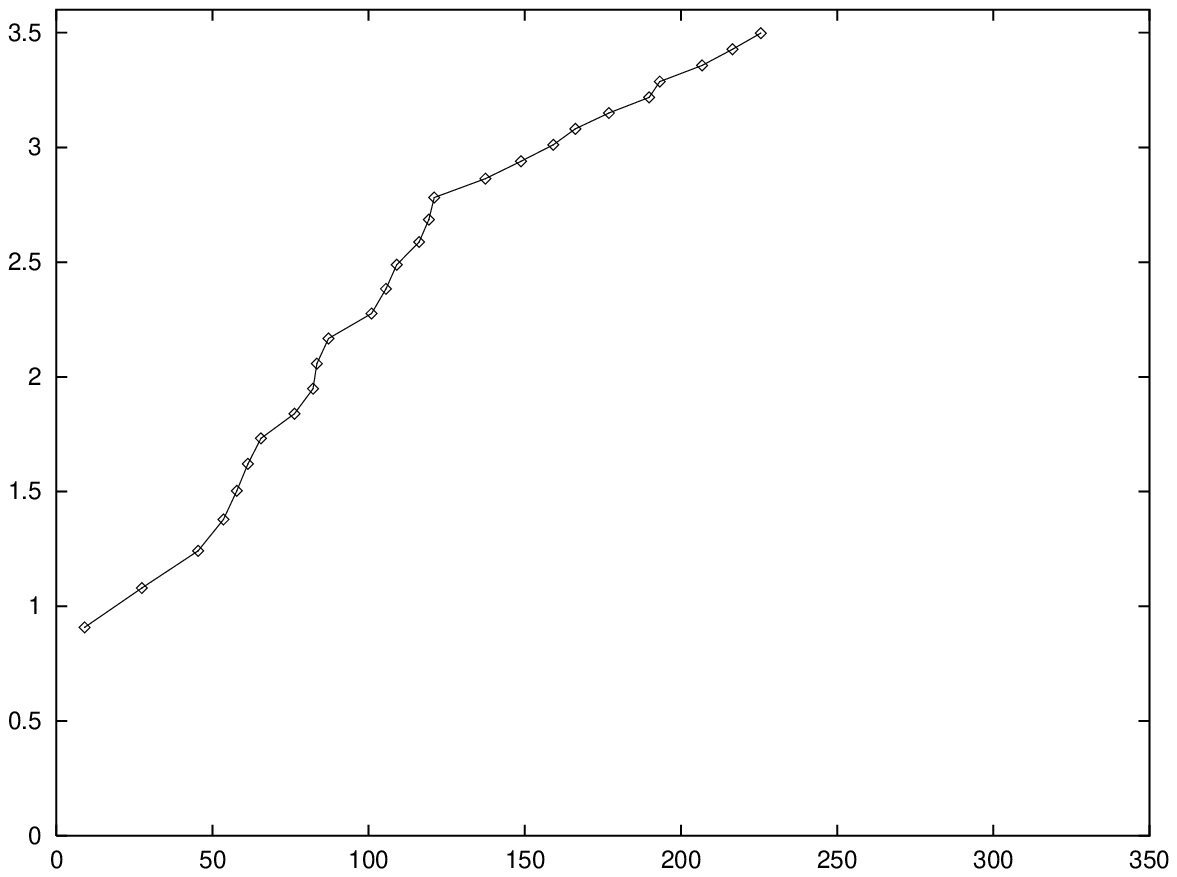,width=7.5cm}
\epsfig{file=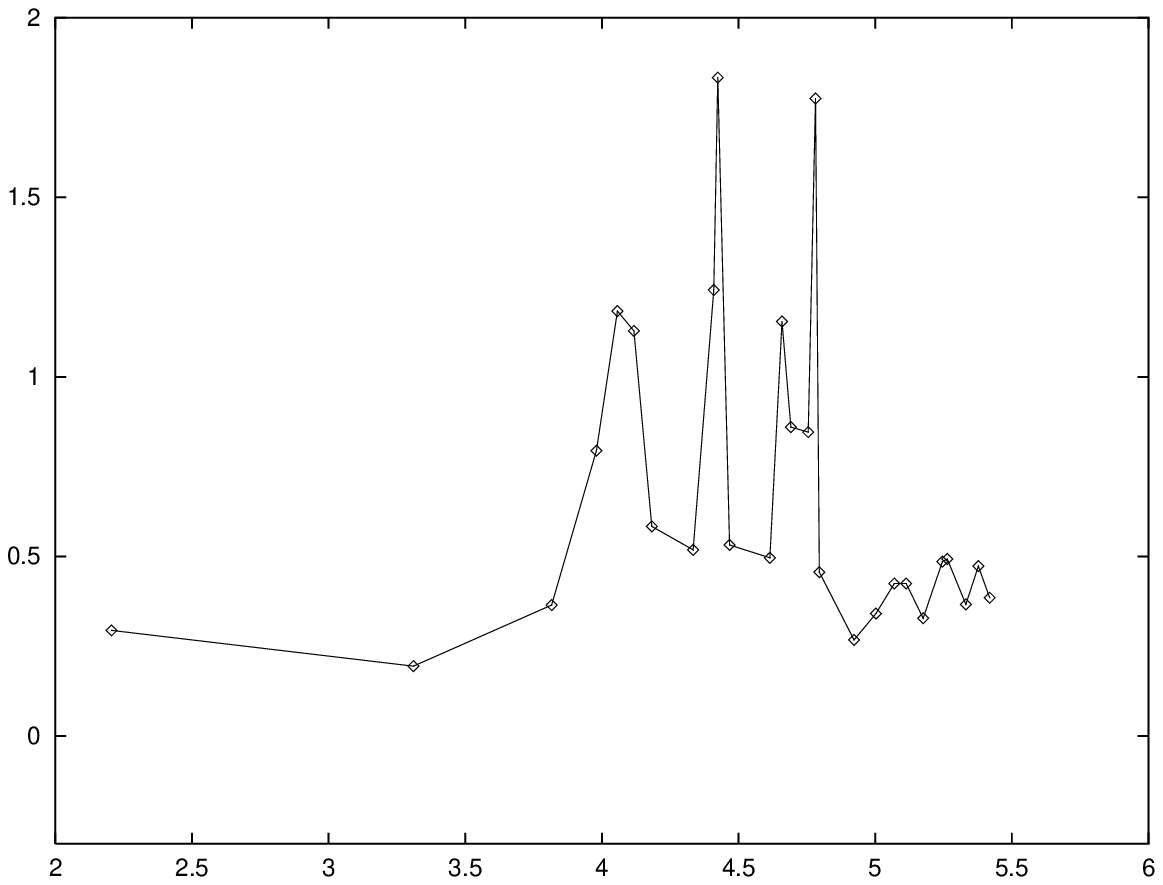,width=7.5cm}
\caption{\protect\label{cumart3} The cumulative distributions of the data set
in figure \protect\ref{art3} and it's logarithmic derivative.}

\vspace{1cm}

\epsfig{file=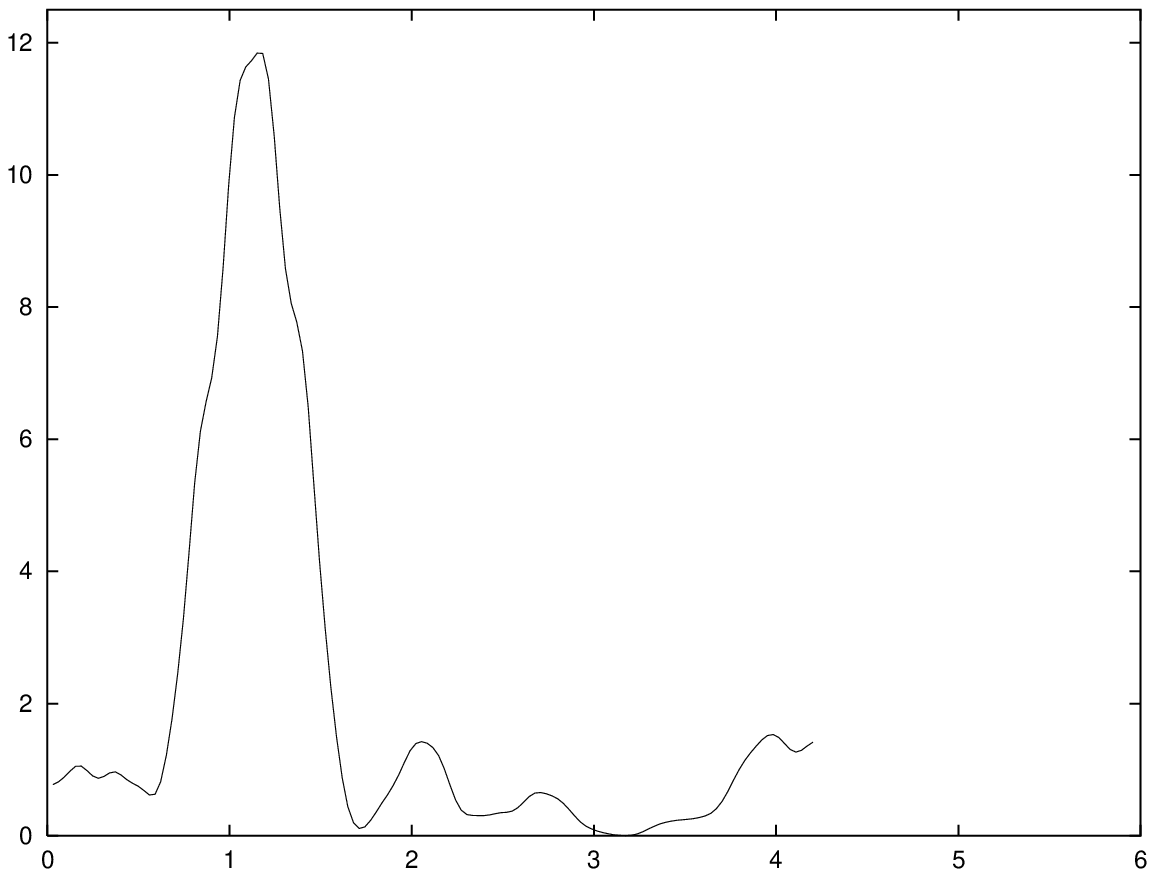,width=7.5cm}
\epsfig{file=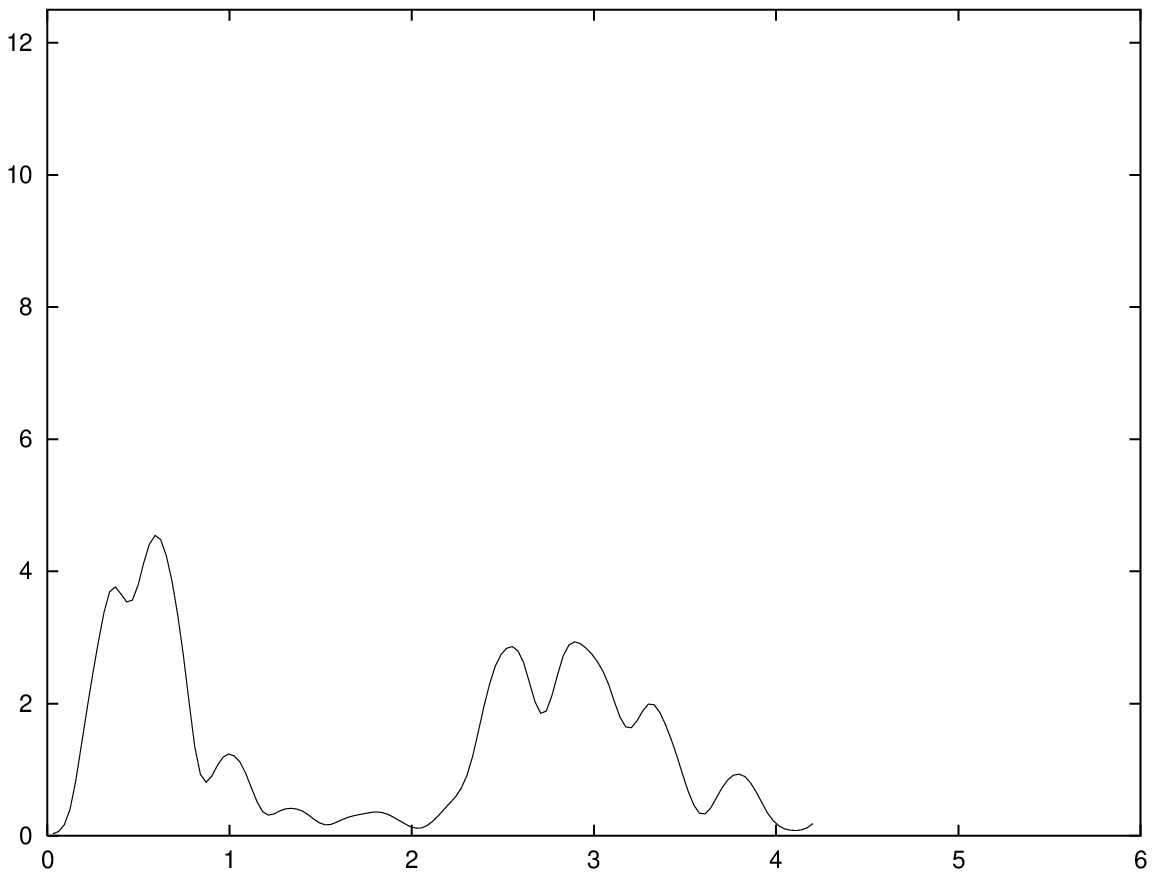,width=7.5cm}
\caption{\protect\label{lombart3}Lomb periodograms of the derivatives in
figures \protect\ref{art3} (left) and \protect\ref{cumart3} (right). }

\end{center}
\end{figure}

Detecting log-periodic structures in noisy, intermittent data is rather
subtle and apparently innocent data manipulations, such as using the
cumulative distribution rather than the data itself, may completely erase such
structures. Furthermore, since these structures are {\it log}-periodic it
is often impossible to obtain a large number of oscillations, since
numerical and experimental data seldom cover more than a few decades. Hence,
great care must be taken not only with respect to what quantity to
analyse in an attempt to quantify the strength and frequency of a log-periodic
signal but also how to average different realisations. Here the question of
noise also enters crucially, since the higher the noise level the more
oscillations are necessary. The next section will try to shed light on these
technical questions and provide guidelines for the analysis of the
turbulent data.

If we suspect that some distribution $p\lp x\rp$ obeys
\be
p\lp x\rp \propto x^\alpha
\ee
and furthermore believe $\alpha$ to be complex, the most direct quantification
is to look at the logarithmic derivative
\be \label{eqderiv}
\frac{d \ln p\lp x\rp}{d\ln x} = \alpha .
\ee

There are a number of ways to estimate the derivative from data, but they
more or less amount to the same thing, namely locally estimating the data by
some function, typically a nth degree polynomial and then performing the
derivative.
Performing the derivative directly on discrete data amounts to drawing a
line between
consecutive data points and using its slope as the derivative. A
more sophisticated Savitsky-Golav filter \cite{NumRec} uses polynomials of
arbitrary degree and preserves higher moments of the data, such as the
variance, which becomes suppressed with a moving average (hence, a moving
average is unsuitable in an analysis for log-periodicity). However, in the
following
we will use the simplest numerical procedure to calculate the derivative, i.e.,
\be
 \frac{d \ln p\lp x\rp}{d \ln x}\left|_{x=x_i} \right. =
\frac{\ln p\lp x_{i+1}\rp - \ln p\lp x_{i-1}\rp}{ \ln x_{i+1} - \ln x_{i-1}}
\ee
in order to keep the numerical manipulation transparent.

A numerical tool that we will use extensively in this paper is the so-called
Lomb periodogram \cite{NumRec}. It corresponds to a harmonic analysis
using a series of local fits of a cosine (with a phase) with some user
chosen range of frequencies. The advantage of the Lomb periodogram over
a Fast Fourier transform is that the points does not have to be equidistantly
sampled, which is the case here.

\subsection{The Lomb periodogram and integration}

A method often used in order to investigate fluctuations in data is to fit
the average behaviour of the data by some function which is then subtracted.
This method can be justified if you {\it only} want to estimate the error of
that specific fit. However, you cannot use it to extract oscillations unless
you have many and if that's the case you can do much better with the data
itself. The problem with the procedure is that the somewhat arbitrary
``truncation'' of
the data set in both ends will produce a bias due to the randomness of the
phases of each realisation.  If some sort of de-trending is to be used
in order to accentuate the more subtle correlations, such as log-periodicity,
then it always makes more sense to {\it divide} the data by the leading trend
rather than subtracting it. However, it's a non-trivial task to get an
un-biased estimate of such a leading trend due to the same arbitrary
truncation. When this is possible, such a de-trending is a very powerful
way of visualising log-periodic oscillations. This has been clearly
illustrated in the case of financial data \cite{risk}.

Another troublesome feature of extracting log-periodic components from data
is that they can easily be destroyed by ``innocent'' data manipulation such
as calculating the cumulative distribution. An example where this is clearly
the case is shown below. We have generated $10$
data sets with $30$ points each using the equation
\be \label{surreq1}
y\lp t\rp = t^{-0.5}\left[ 1+ 0.1\cdot\cos\lp 2\pi f\ln t \rp
\right]~~~~~{\rm with}~~f =1.114
\ee
where the sampling was random in the sense that the spacing between two
consecutive points was chosen from the interval $\left[ 0:20\right]$ with
uniform probability.

Figures (\ref{art1},\ref{art2},\ref{art3}) show three
typical realizations of the random sampling of (\ref{surreq1}) together
with the
corresponding discrete estimations of their logarithmic derivatives.
Figures (\ref{cumart1},\ref{cumart2},\ref{cumart3}) show the corresponding
integrals (cumulatives) of the three samples together with
the discrete estimations of their logarithmic derivatives.
Figures (\ref{lombart1},\ref{lombart2},\ref{lombart3}) compare the
Lomb spectrum of the direct and cumulative sampling presented in the
previous figures.
Whereas we clearly extract the log-periodic component
in the signal and its frequency $f=1.1$ with a level of significance better
than
$0.995$ (see Ref.\cite{NumRec} for the quantification of the significance
from the peak level), the signal has completely disappeared in the
cumulative distribution.

This not only
shows that a relatively strong ($10\%$ in amplitude)
log-periodic signal will be destroyed by an seemingly
innocent summation but also that without noise only $\approx$ 2
oscillations is enough to qualify log-periodicity.
Similarly to the performance of a FFT which depends on the
 ratio of the measured frequency over the Nyquist frequency as
well as the length of the time series, the performance of the Lomb periodogram
depends on the number of points per oscillation and the number of oscillations.
In fact, this example
confirms very well the general belief that noise in data can be ``integrated
out'' and that constructing cumulatives
provides a very efficient low-pass ($1/f^2$) filter. However, for our present
task of identifying log-periodicity, it also shows the danger of this
procedure which integrates out both the noise and the useful signal.

\subsection{Additive noise}

\begin{figure}
\begin{center}
\epsfig{file=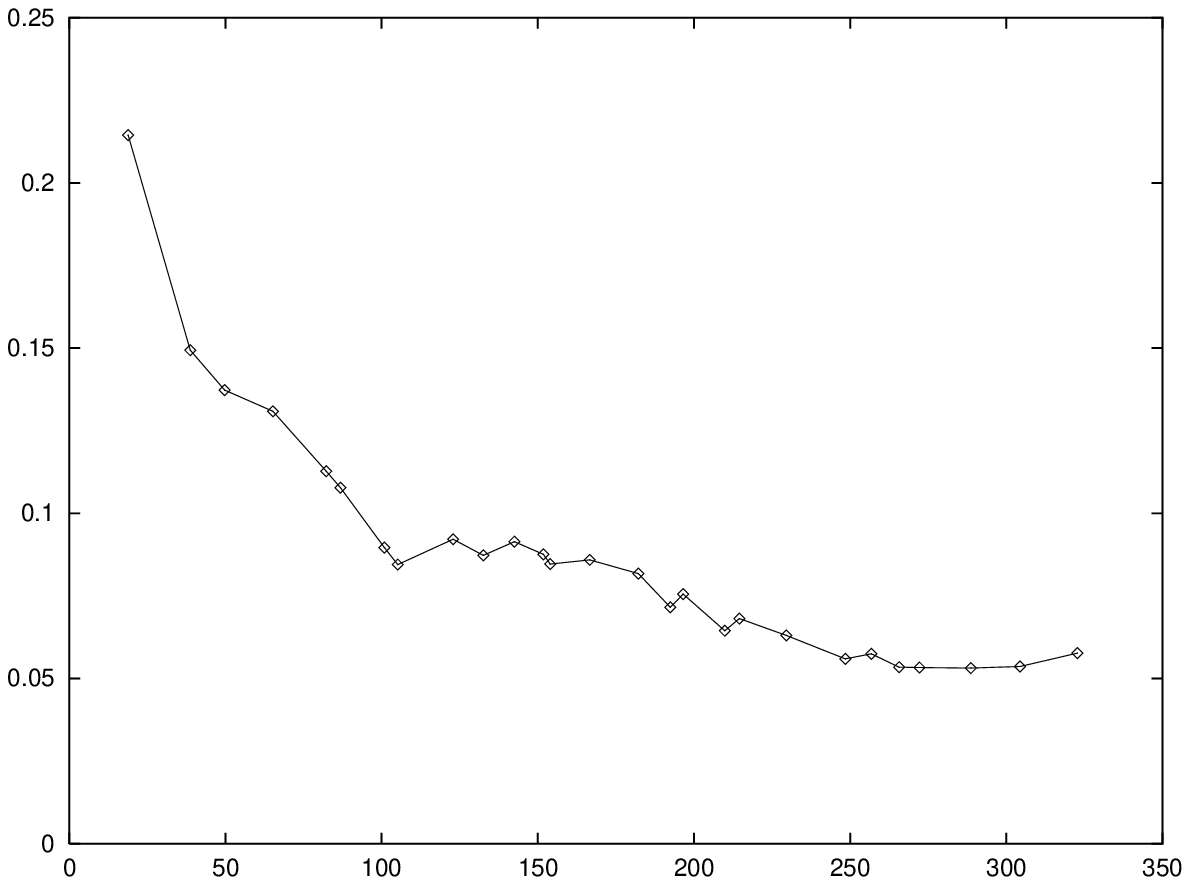,width=7.5cm}
\epsfig{file=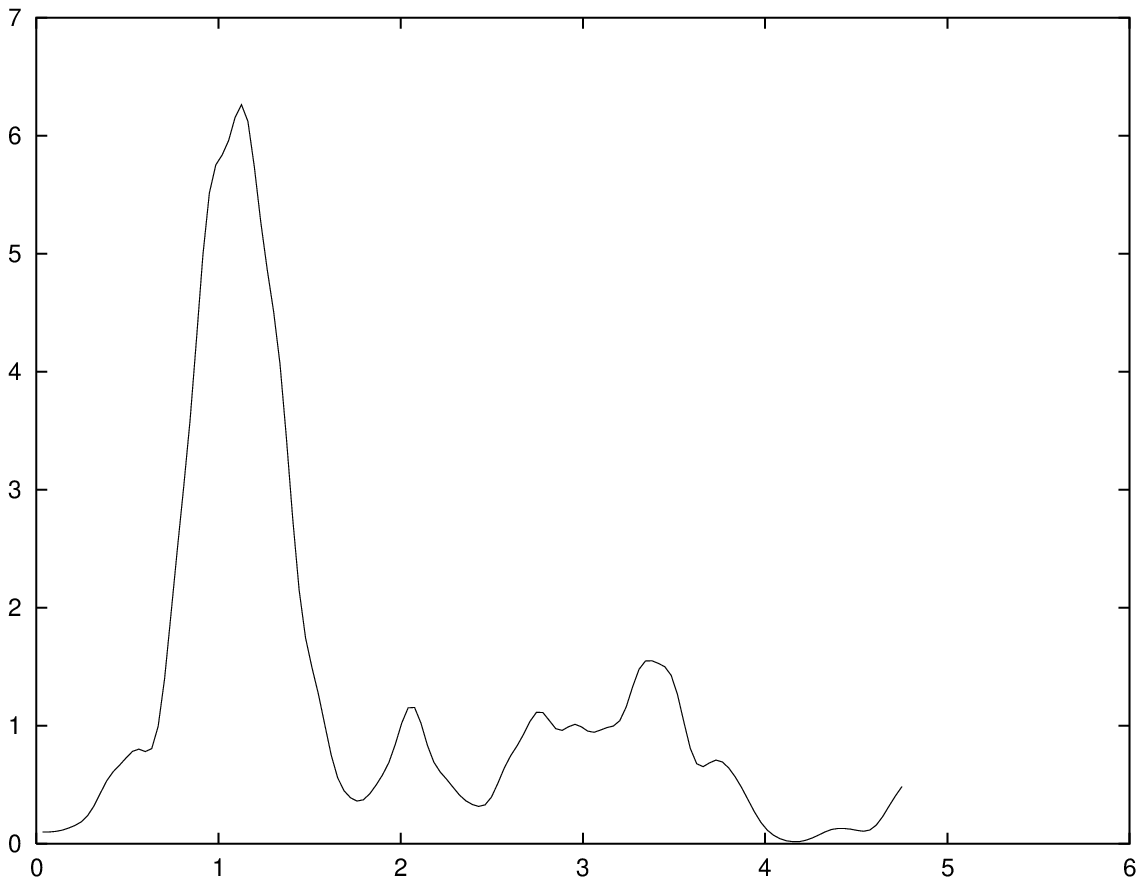,width=7.5cm}
\caption{\protect\label{noiart1} First noisy data set (left) and Lomb
periodogram of its logarithmic derivative (right).}

\vspace{1cm}

\epsfig{file=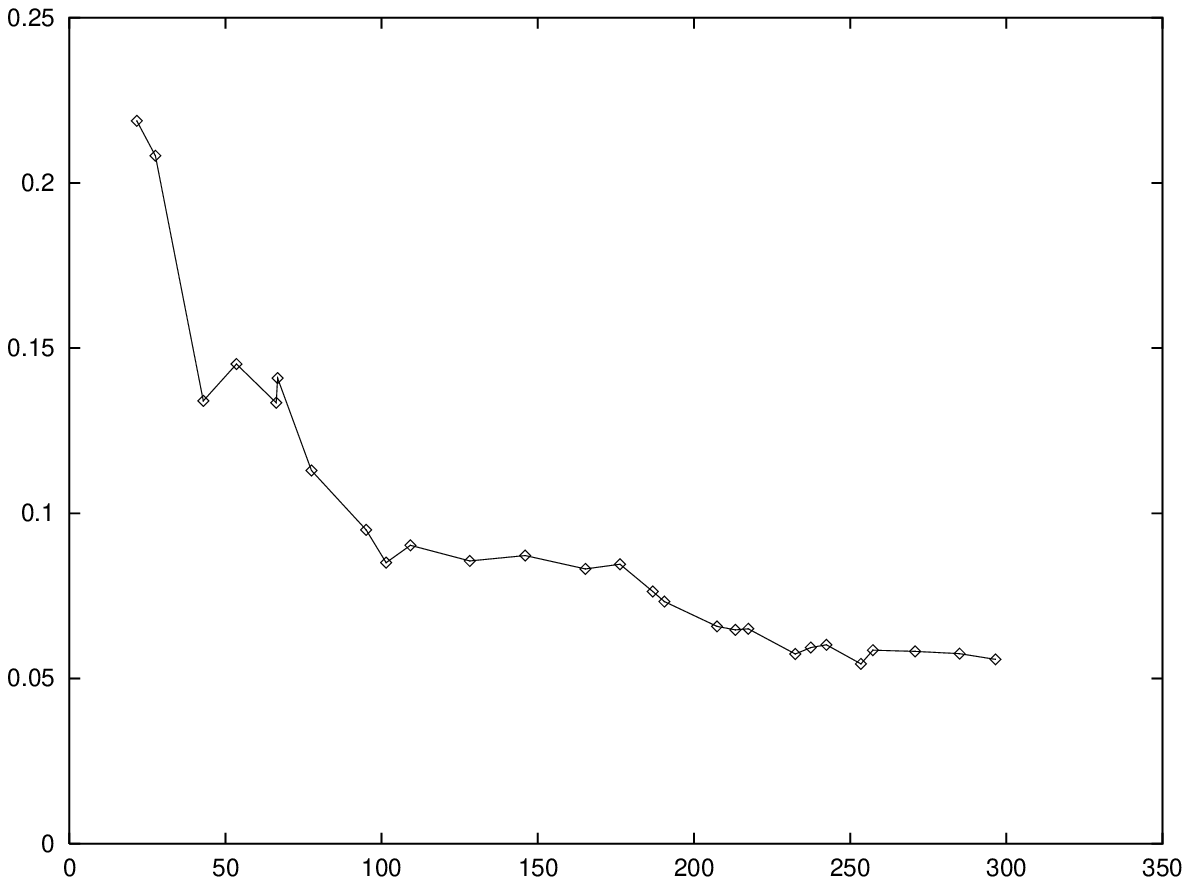,width=7.5cm}
\epsfig{file=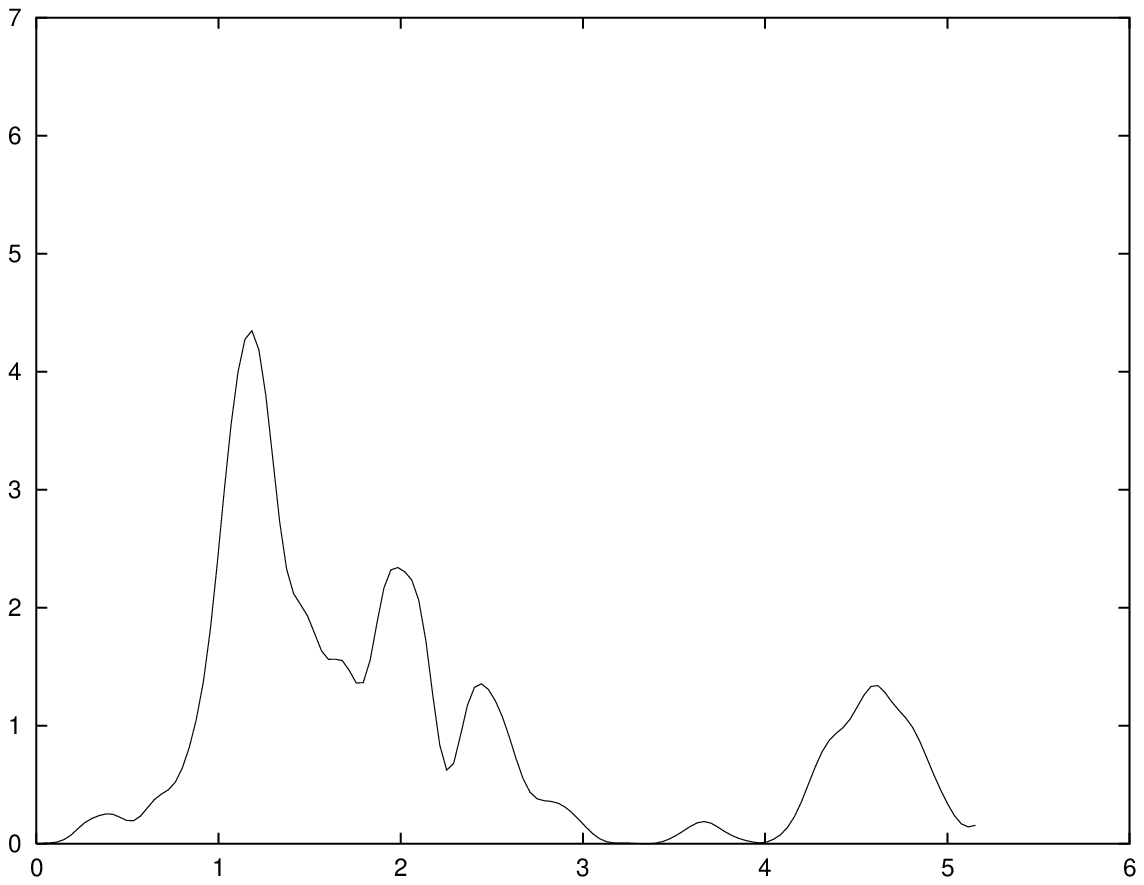,width=7.5cm}
\caption{\protect\label{noiart2} Second noisy data set (left) and Lomb
periodogram of its logarithmic derivative (right).}

\vspace{1cm}

\epsfig{file=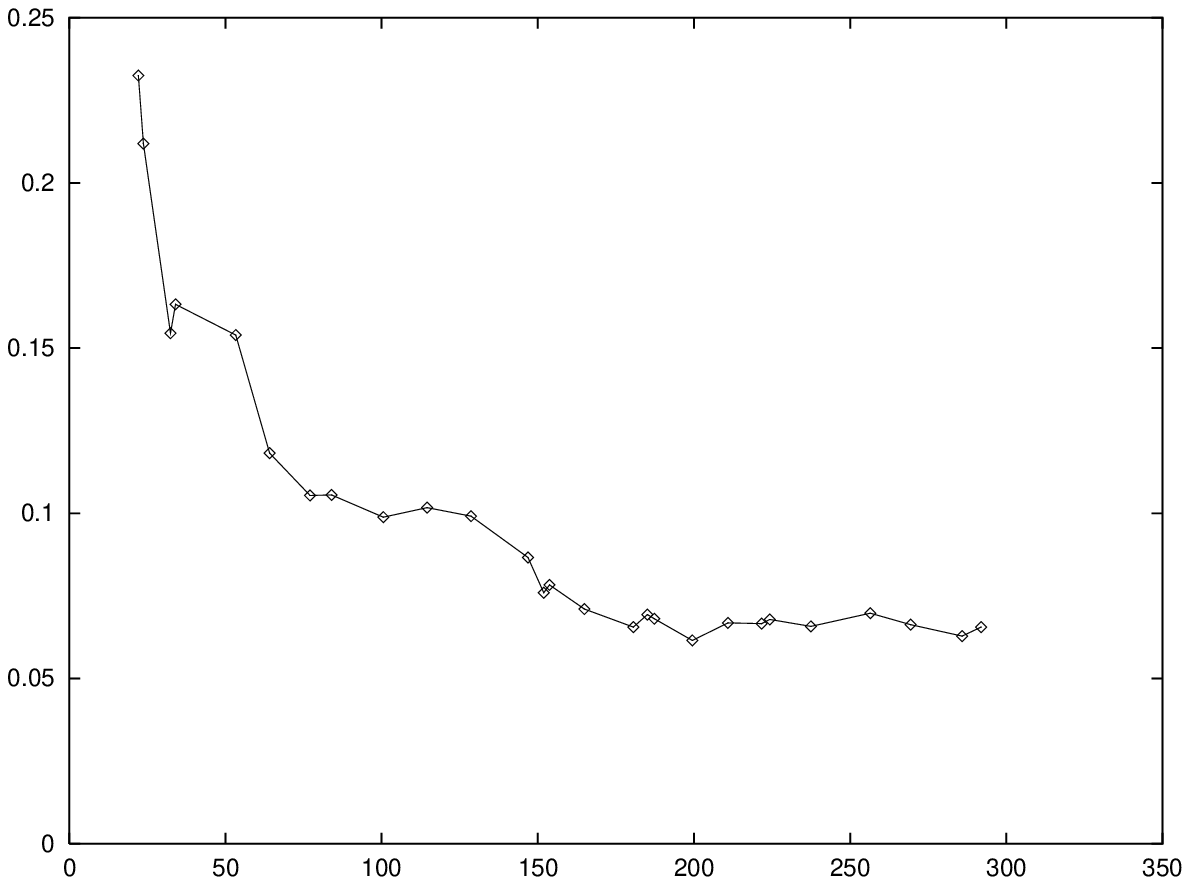,width=7.5cm}
\epsfig{file=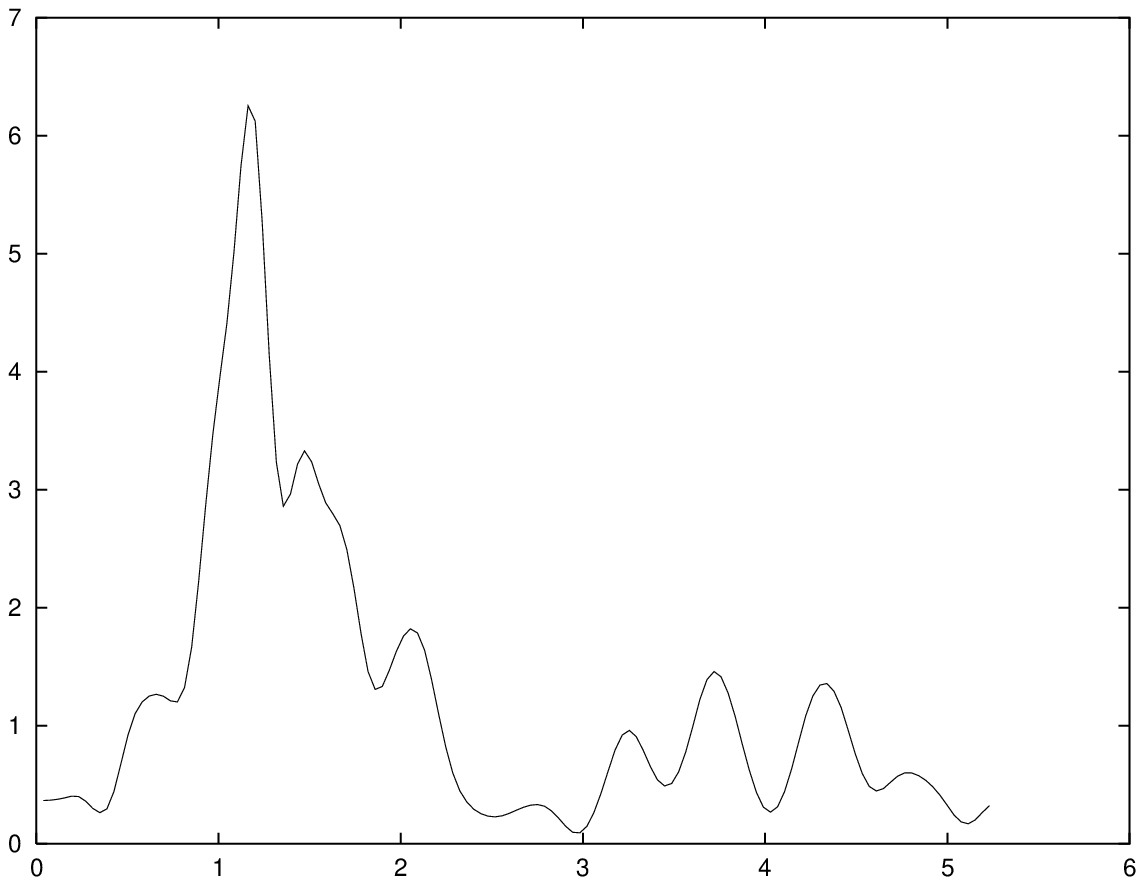,width=7.5cm}
\caption{\protect\label{noiart3} Third noisy data set(left) and Lomb
periodogram of its logarithmic derivative (right).}
\end{center}
\end{figure}

The process used in
generating figures (\ref{art1},\ref{art2},\ref{art3}) is already
noisy in some sense due to the random sampling.\footnote{It is probably better
to keep the vocable ``noise'' for processes that deteriorate the signal.
In the case of a random
sampling of $n$ points in some interval $T$, frequency components which are
larger than the Nyquist frequency of the corresponding evenly sampled points
$\frac{N}{2T}$ can be identified. Hence, the random sampling is in fact
improving the signal-to-noise ratio for high frequencies.}
We now investigate the effect of additive noise, in addition to the irregular
sampling. We have thus added a noise term to equation (\ref{surreq1})
and rewritten it as
\be \label{surreq2}
y\lp t\rp =(1+k(0.5-ran))t^{-0.5}\left[ 1+ 0.1\cdot\cos\lp 2\pi f\ln t \rp
\right] ~,
~~~~{\rm with}~~f =1.114~,
\ee
where $k$ is the amplitude of the noise and $ran$ is the intrinsic random
number generator of Fortran. In figures \ref{noiart1} to \ref{noiart3} we see 3
examples of such synthetic data sets and their corresponding Lomb
periodograms. Here $k=0.1$ is used corresponding to a noise level half of
the amplitude of the log-periodic oscillations. What we see now is that the
confidence interval has come down considerable, but the peak is still
correctly positioned for all 10 data sets of which 3 arbitrary sets are
shown here. In fact, what we see here is not very much different from the
analysis of diffusion limited aggregation (DLA)
presented in \cite{dla,thesis}. The individual data set do
not carry
a very large significance, but the consistence of the analysis for the
entire ensemble of data sets do.

If we increase the noise by a factor of $2$,
{\it i.e.}, to the same amplitude as that of the log-periodic oscillation
things become very murky, but it is still possible to identify the
log-periodic oscillations using the results for the entire {\it ensemble}.
If we, as for DLA \cite{dla,thesis}, record the two highest peaks for the
$10$ data sets, we get the distribution shown in figure \ref{histo}. It is
rather remarkably that, with the Lomb periodogram, we can retrieve
log-periodicity with such quality in data which has a noise amplitude of
the same
size as that of the log-periodicity. In the next section, we shall see a
more elegant way of retrieving log-periodicity in noisy data.

\subsection{The method of canonical averaging}\label{cano}

The method of ``canonical averaging'' has been introduced \cite{Pazmandi}
to reveal subtle correlations which would disappear in the usual
ensemble averaging. We have tested a specific implementation of it on
log-periodicity
in DLA clusters and rupture time-to-failure processes \cite{dla,thesis,fuse},
based
on a sample-dependent phase determined from the maximum of a
sample-dependent susceptibility.

Recall that the problem
comes from the fact that the argument
$2\pi f\ln t$ in the cosine in equation (\ref{surreq1}) is defined up to a
phase, since adding a phase simply corresponds to a change of time units.
In other words,
the universal physical quantity is the log-frequency
$f$ while the phase is expected to be sample-specific
and to fluctuate from realizations to realizations \cite{fuse}. If this problem
is not addressed properly, standard averaging procedure mix measurements with
different phases leading to a destructive interference of the log-periodic
oscillations.

\begin{figure}
\begin{center}
\epsfig{file=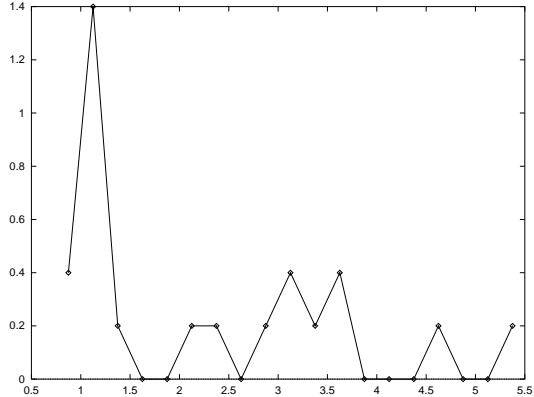,width=7.5cm}
\caption{\protect\label{histo} Histogram of the frequency distribution for the
two most significant peaks in the Lomb periodogram of $10$ data sets with a
noise amplitude of the same size as the amplitude of the log-periodic
oscillations. The peak in the histogram correspond to a frequency of
$\approx 1.12$ in very good agreement with that of $f \approx 1.114$
used.}
\end{center}
\end{figure}

\begin{figure}
\begin{center}
\epsfig{file=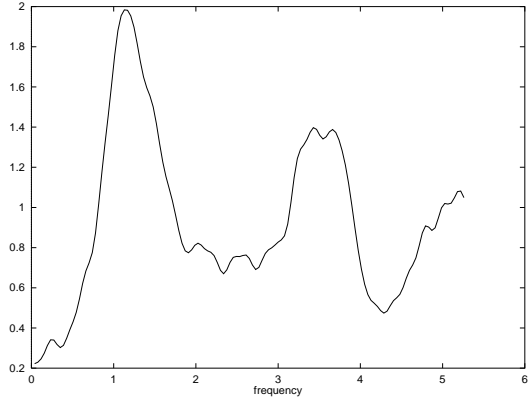,width=7.5cm}
\caption{\protect\label{avelomb} The averaged Lomb periodograms of the 10 data
sets used in the histogram shown in figure \protect\ref{histo}. We emphasise
that the amplitude of the noise was of the same size as the amplitude of the
log-periodic oscillations. Note that the true confidence interval does not
correspond to the values shown, since it's an average over 10 data sets. The
value of the two peaks are $1.13$ and $3.4$ and in excellent agreement with
the 1. and 3. harmonic of the frequency $f \approx 1.114$ used in generating
the data sets.}
\end{center}
\end{figure}

We now propose a much simpler method than previously used \cite{fuse} whose
principle is to ``re-phase'' the oscillations. The method is based on the
fact that a Lomb periodogram is a simple
cosine fit, while the phase is not playing any role. Hence,
the periodogram corresponds to a removal of the phase from the data. This
means that averaging the periodograms of different data sets is exactly the
``re-phasing'' we are looking for. In figure \ref{avelomb}, we show the
averaged
periodogram for the 10 data sets used in the histogram shown in figure
\ref{histo}. The two peaks corresponds two frequencies of and $1.13$ and
$3.4$, respectively. The first and most significant peak is in very good
agreement with the frequency of $1.114$ used in generating the data sets.
The second peak is furthermore in excellent agreement with that of the third
harmonic. We remind the reader that figure
\ref{avelomb} comes from averaging periodograms of noisy data with a
noise-amplitude of the {\it same size} as the amplitude of the log-periodic
oscillations. Nevertheless, the log-frequency component of $\approx 1.1$
present in the data is unambiguously extracted as well as its third harmonic
hardly visible in the histogram in figure \ref{histo}. This illustrates not
only the efficiency of the Lomb periodogram in retrieving periodic trends, but
also that of the canonical averaging scheme for re-phasing the oscillations.

This concludes the numerical tests of the tools to be used
in the experimental data analysis that
we present in the next section.

\section{Experiments on 2-d freely decaying turbulence}

The experimental data analysed here are some of those presented in
\cite{Hansen} (Fig.'s 6 and 7). We will briefly describe the experimental
techniques and the results that are used in the present context.

\subsection{Experimental setup}

The  flow  is generated in a thin, density  stratified layer
of electrolyte, using an electro-magnetic forcing.  During a short initial
forcing  period, an 8 $\times$ 8 array of vortices is created, with nearest
neighbours counter-rotating. At a time defined as $t$ = $0$, the forcing
is stopped, and the system is then decaying freely. After a vertical
reorganisation of the flow with a maximum duration of 2 seconds, the
system can be regarded as two-dimensional, except for the effect of
bottom friction \cite{Paret97}, which causes an exponential decay of the
total energy of the system with a well-defined time constant.
A rescaling of time
is applied under which we can consider the system as truly two-dimensional
\cite{Hansen}.

The size of the experimental system is 15 cm $\times$ 15 cm, and the
initial Reynolds number is typically 1800.
The duration of the experiments is approximately 12 seconds in  rescaled
time units.

\subsection{Determination of vortex statistics}

The velocity field ${\bf v}(x,y)$ on the free surface of the fluid is
determined using the Particle Image Velocimetry (PIV) technique described
in \cite{Cardoso94}, with a resolution of $40 \times 40$ measurement points.
From the velocity field, the vorticity field
$\omega (x,y)  = \partial_x v_y - \partial_y v_x$ is computed using a
polynomial fit. Typically 70 velocity fields are calculated during the
time of the experiment.
The vorticity field is analysed for vortices by
searching for local extrema with the absolute vorticity above a
threshold. The results are not sensitive to the choice of threshold,
and the recognition method has been checked with other methods.
Thus we can define a number $n$ of vortices for each velocity field.
The position
of a vortex is defined as the position of the vorticity extremum.
This is used to extract the mean nearest-neighbour separation $r$
of the vortices. The size of a vortex is defined by
the connected area with an absolute vorticity above the threshold.
For each velocity field the mean vortex radius $a$ is
computed from the mean vortex size.

\subsection{Scaling laws}

Rapidly after the forcing has been stopped, like-sign vortices start
to merge. Fewer and larger structures are thus formed, and this
continues until the energy of the system is so small that no further
evolution of the vortices can be observed.

We presented \cite{Hansen} measurements that are averages of
9 experimental realizations (in the case of vortex radius, only 7 realizations
were used).
We found that for the time period from $t$ = $1.5$ s to $t$ = $10$ s
the quantities defined above follow the power laws
\begin{eqnarray}
\label{geomQ}
n(t) & \sim & t^{-0.70 \pm 0.1} \nonumber \\
a(t) & \sim & t^{0.21 \pm 0.06} \\
r(t) & \sim & t^{0.38 \pm 0.08}. \nonumber
\end{eqnarray}
The error bars are estimated by considering the fluctuations observed
between individual realizations, {\it however} without any re-phasing
of the log-periodic oscillations to be extracted in section \ref{datanaly}.
We note that the average values for the exponents in (\ref{geomQ}) agree
well with the scaling approach \cite{Carnevale91}.

\subsection{Log-periodic analysis of the experimental data} \label{datanaly}

As explained in the previous section, three experimentally independent
quantities has been measured as a function of time: the number of vortices,
$n\lp t \rp$, the mean vortex radius $a\lp t \rp$ and the mean distance
between vortices $r\lp t \rp$. For each of these quantities, the log-periodic
signatures present in the data will be extracted using the tools described
in the section \ref{tools}. Specifically, the logarithmic derivative as
defined by eq. \ref{eqderiv} was calculated from the experimental data.
Truncating the times series in order to avoid end-effects gave $\approx 35$
points for each periodogram to use. This corresponds well to the $30$ points
used in the synthetic data analysis of section \ref{tools}.

\subsubsection{Number of vortices}

\begin{figure}
\begin{center}
\epsfig{file=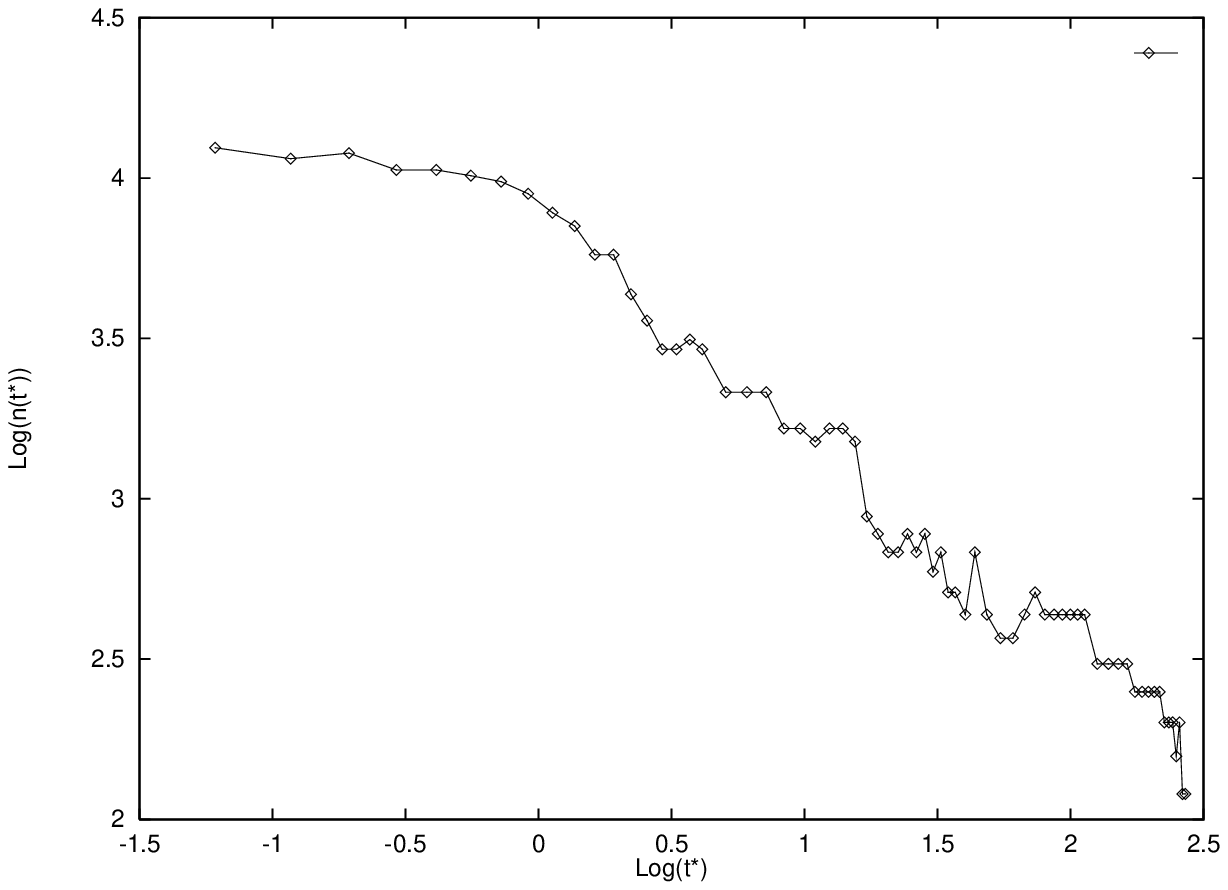,width=7.5cm}
\epsfig{file=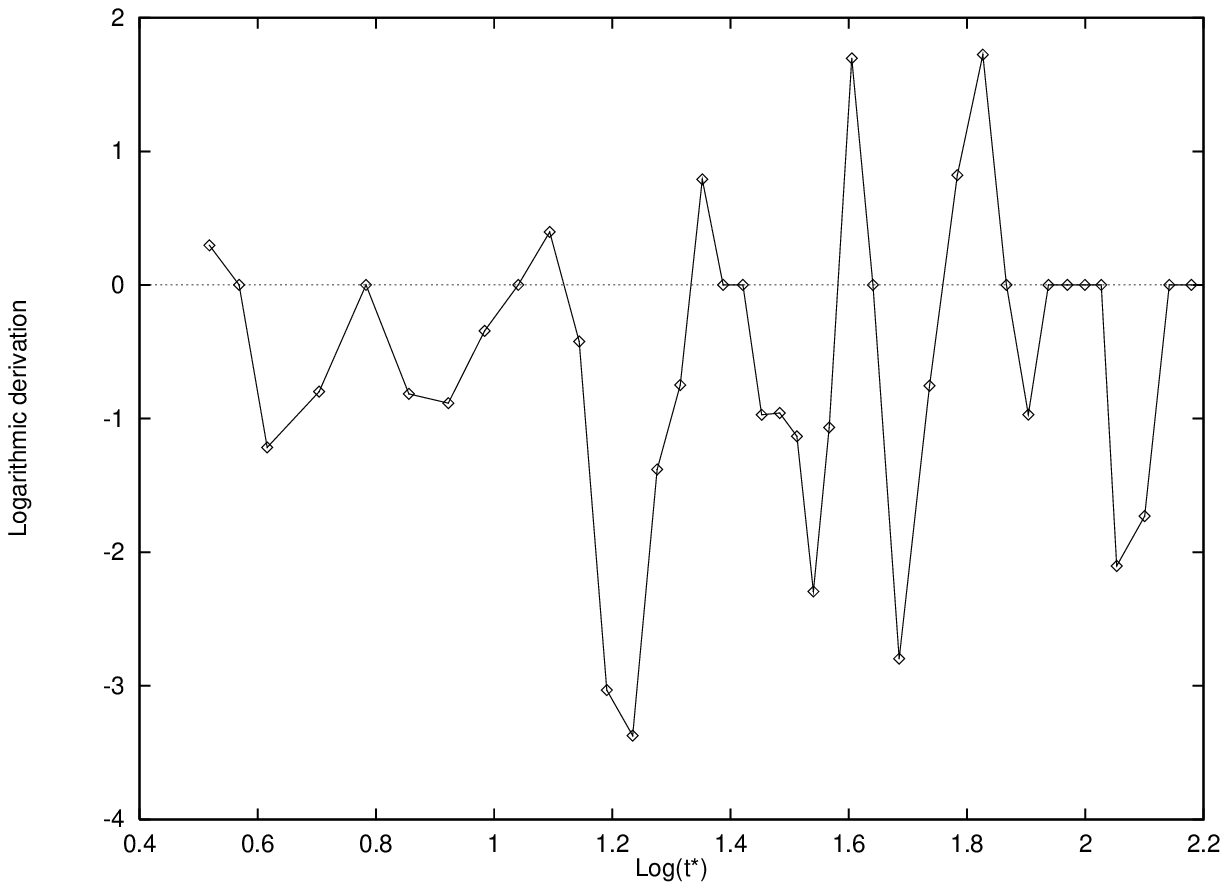,width=7.5cm}
\caption{\protect\label{ndat} Logarithm of the number of vortices
$\log\lp n\lp t\rp \rp$ as a function of the logarithm of re-scaled time
$\log\lp t\rp$ (left). The corresponding logarithmic derivative as a
function of $\log\lp t\rp$ (right).}

\vspace{1cm}

\epsfig{file=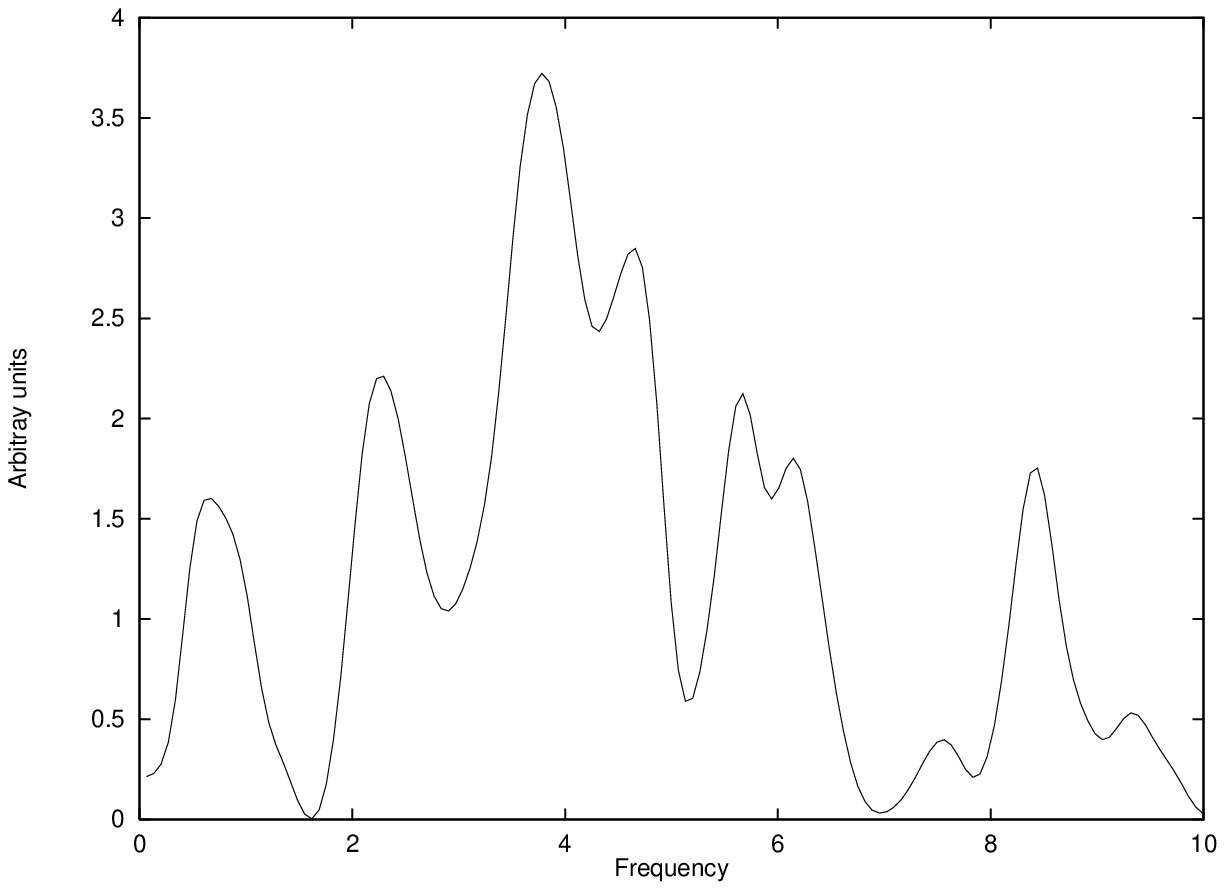,width=7.5cm}
\epsfig{file=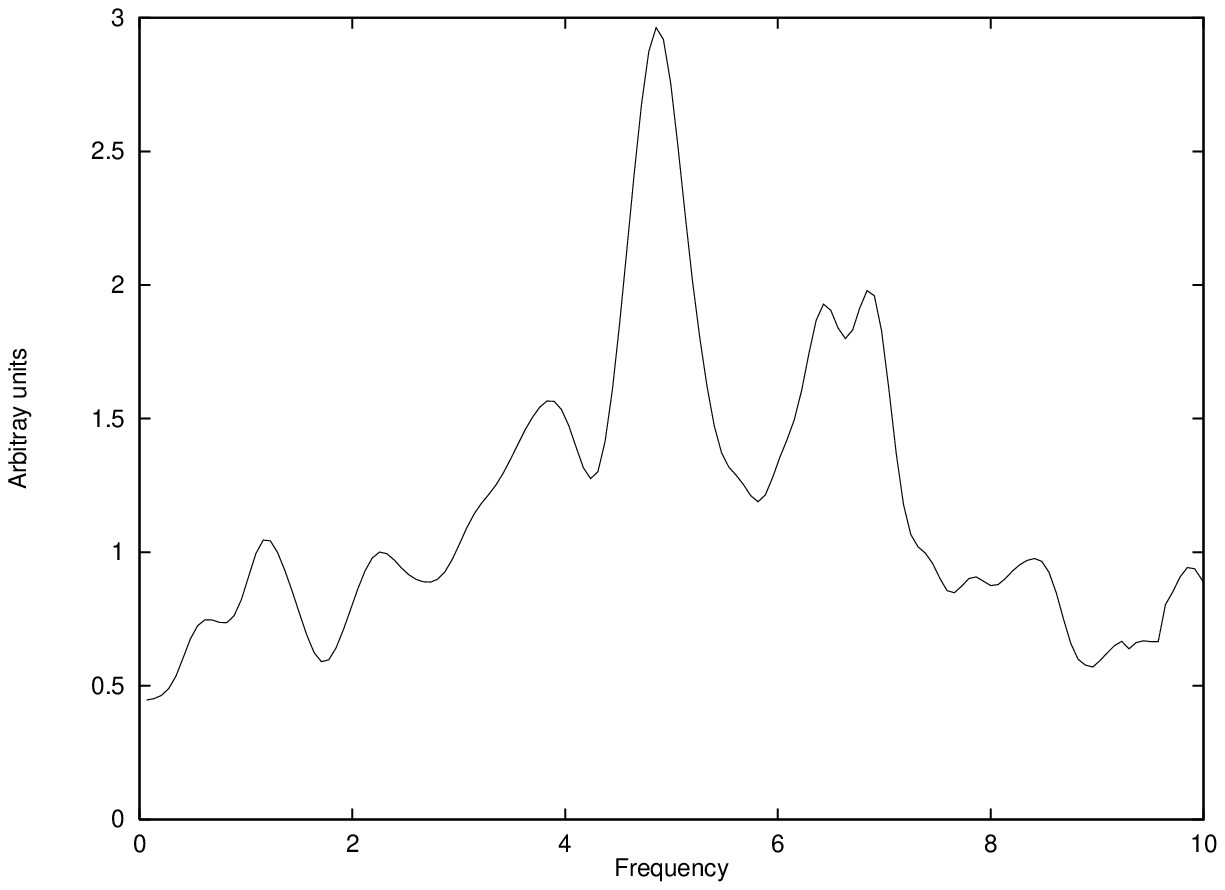,width=7.5cm}
\caption{\protect\label{nfp} Lomb periodogram of the logarithmic
derivative shown in figure \protect\ref{ndat} (left). Lomb periodogram
of the logarithmic derivative of the number of vortices $n\lp t \rp$
as a function of re-scaled time $t$ {\it averaged} over 9 experimental
realisations (right).}
\end{center}
\end{figure}

In figure \ref{ndat}, we see the number of vortices $n(t)$ as a function of
re-scaled time $t$ (left) and its logarithmic derivative (right) for a
specific experimental realisation. In figure \ref{nfp} (left), we see the
corresponding periodogram of the logarithmic derivative shown in figure
\ref{ndat} with a peak at a log-frequency of $\approx 4$. The height
of the peak is only approximately twice the height of the second largest
which means that the confidence of the peak is not very high. Nevertheless,
performing a simple average of the Lomb periodograms for the 9
available experimental
realisations gives us the picture shown in figure \ref{nfp} (right). The
peak has moved up slightly but is still clearly visible, which means that
the position of a peak at a frequency of $\approx 4$ for the single
realisation was not accidental, since a pure noise signal would have
disappeared in an average over 9 realisations. We stress that a frequency of
$\approx 4$ is approximately a factor 5 below the average sampling frequency
of $1.6/35 \approx 22$. This means that the number of points per
oscillation is smaller than that of the synthetic data which explains the
lesser performance.

\subsubsection{Mean separation of vortices}

\begin{figure}
\begin{center}
\epsfig{file=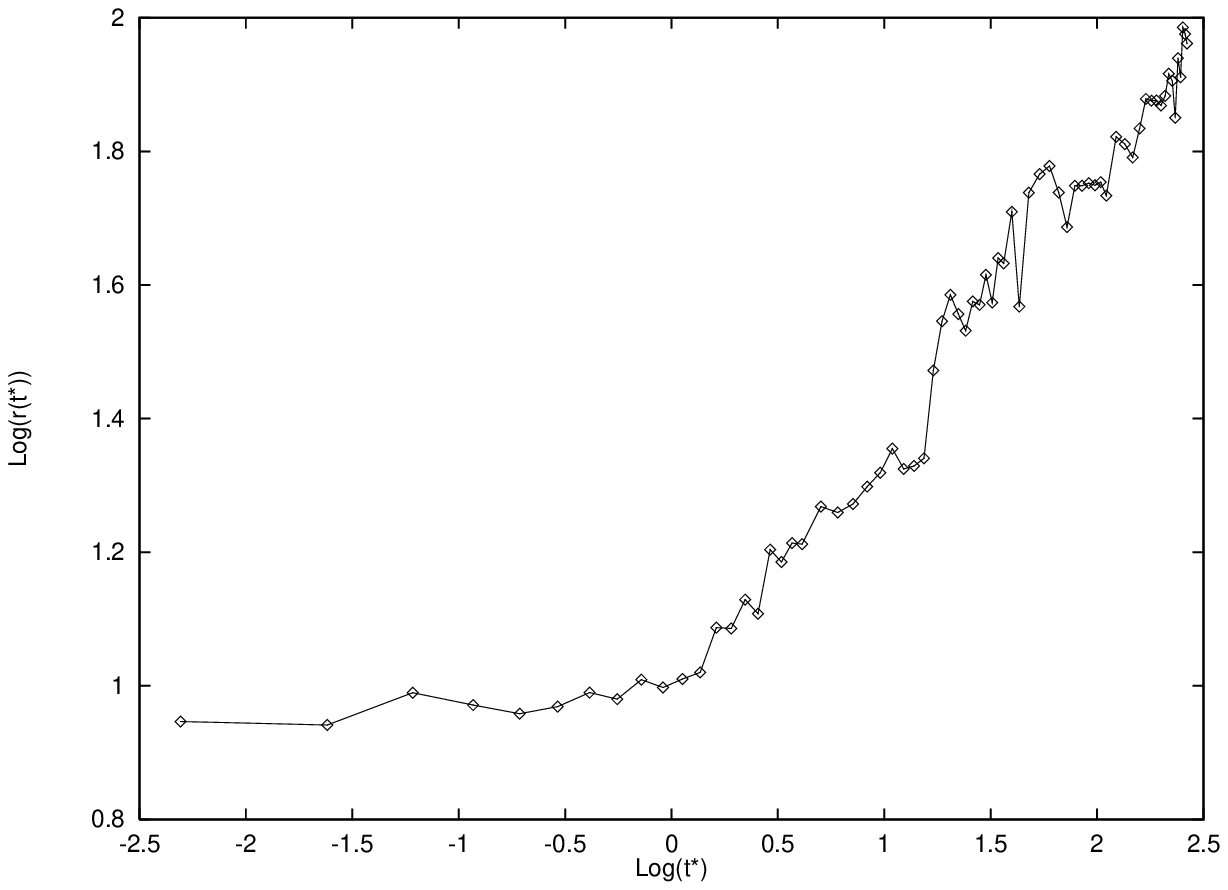,width=7.5cm}
\epsfig{file=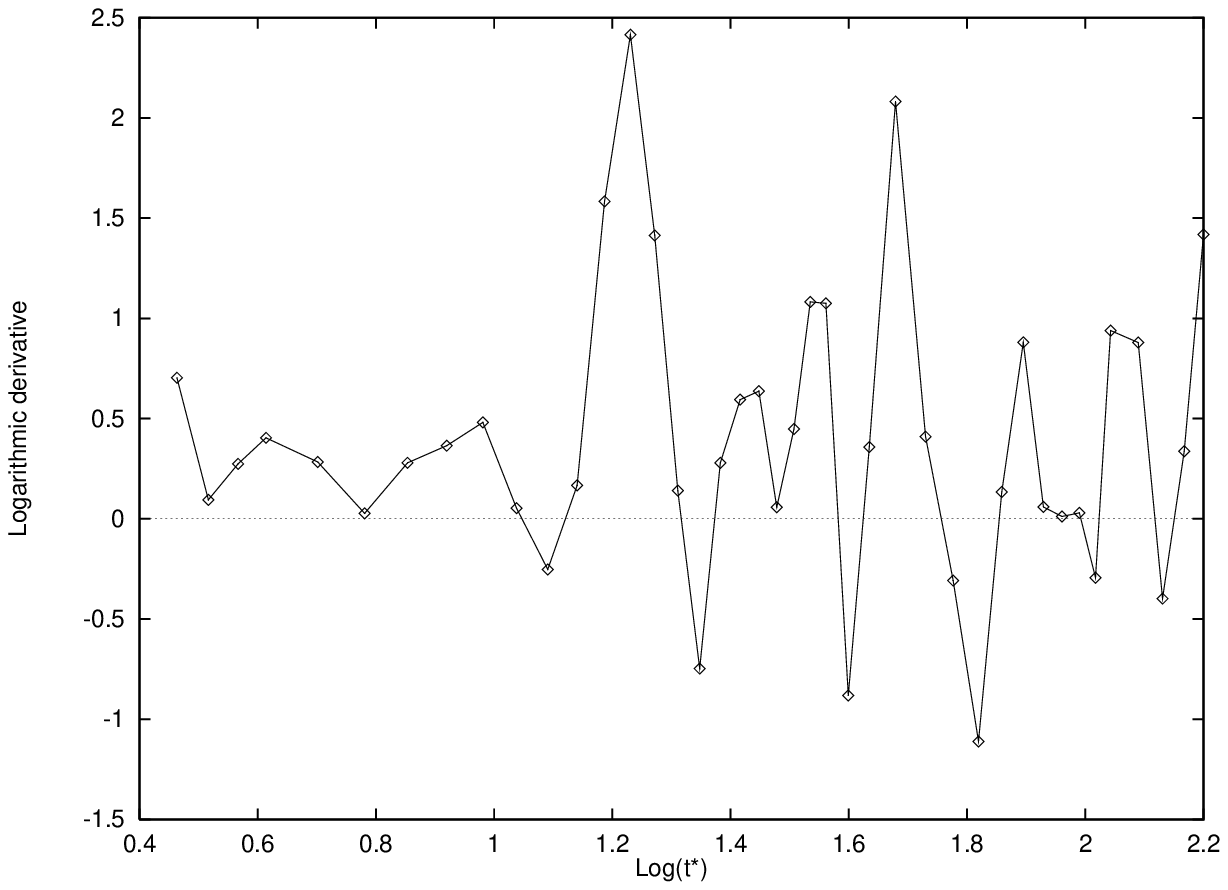,width=7.5cm}
\caption{\protect\label{rdat} Logarithm of the mean separation of
vortices $\log\lp r\lp t\rp \rp$ as a function of the logarithm of
re-scaled time $\log\lp t\rp$ (left). The corresponding logarithmic
derivative as a function of $\log\lp t\rp$ (right).}

\vspace{1cm}

\epsfig{file=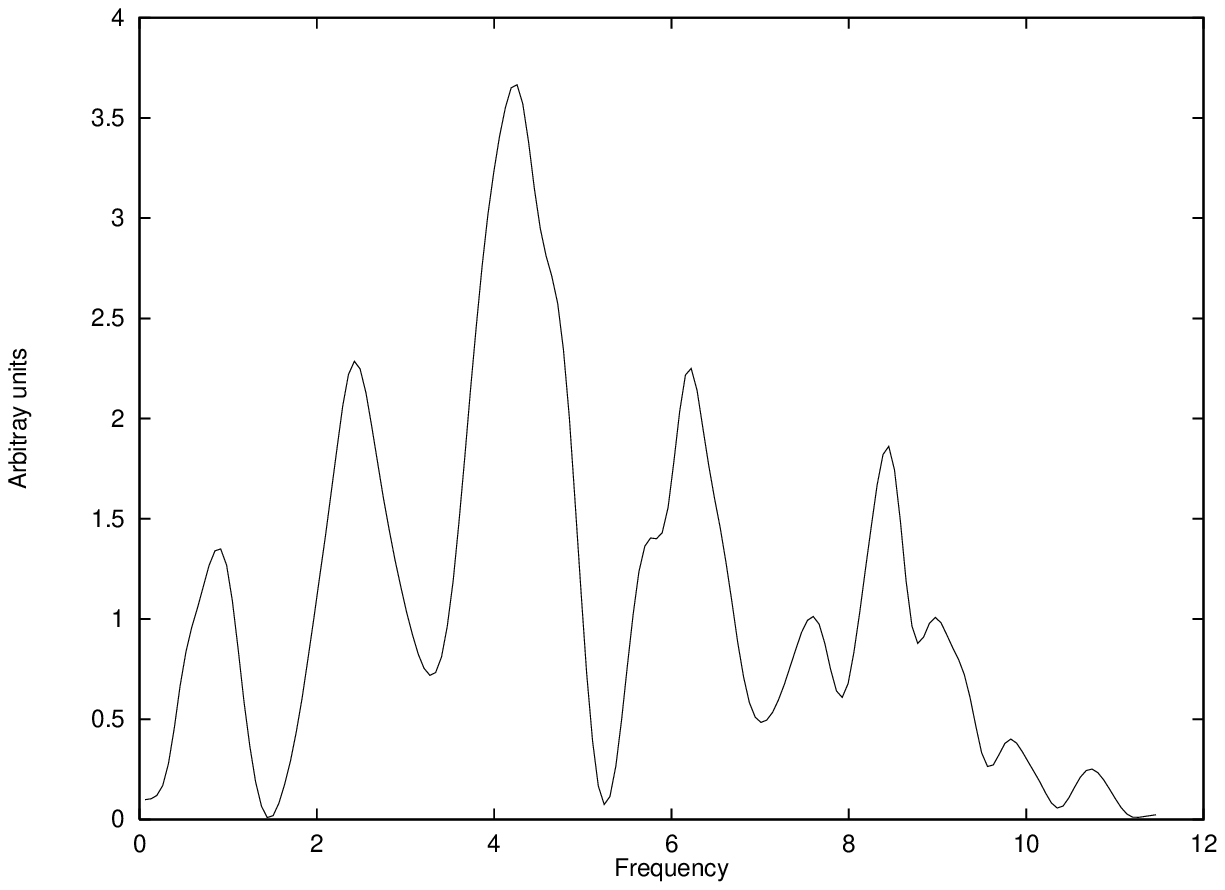,width=7.5cm}
\epsfig{file=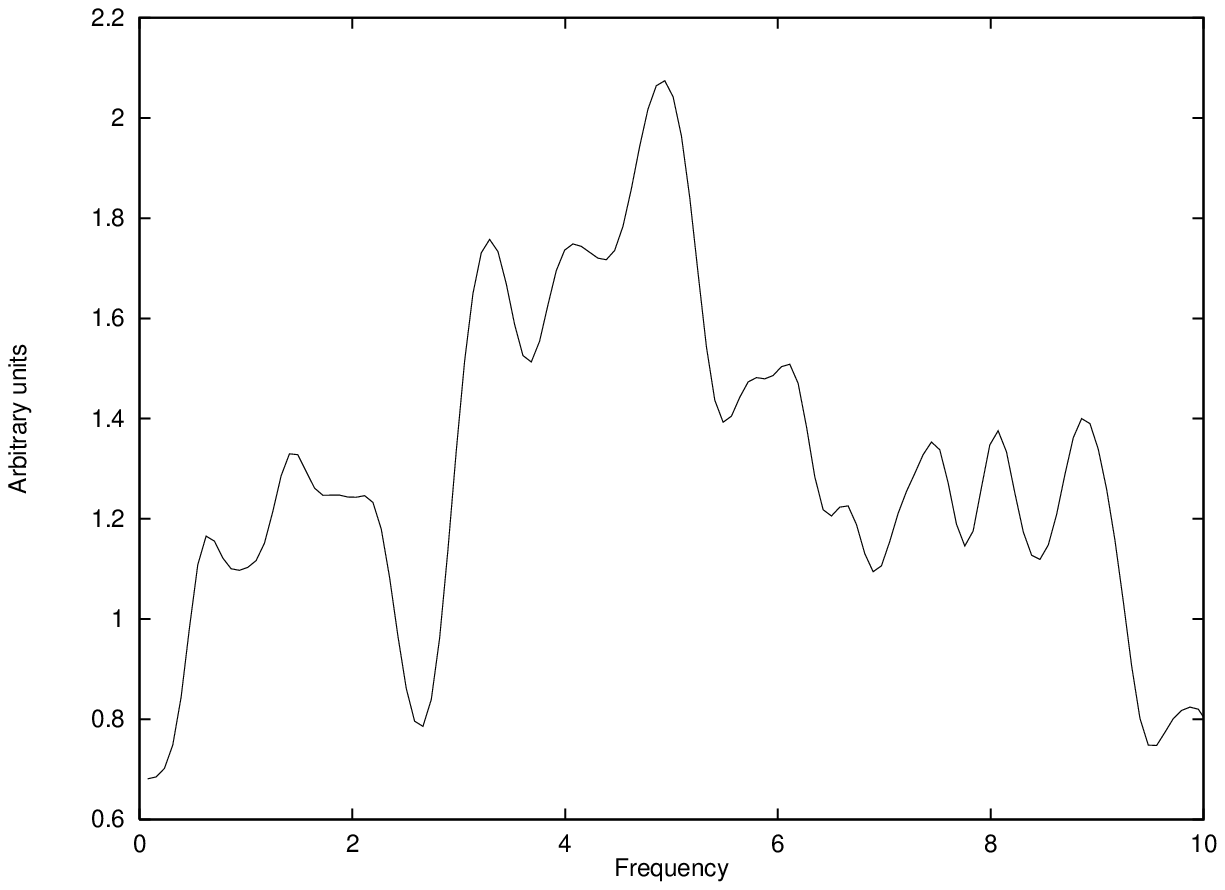,width=7.5cm}
\caption{\protect\label{rfp} Lomb periodogram of the logarithmic
derivative shown in figure \protect\ref{rdat} (left). Lomb periodogram
of the logarithmic derivative of the mean separation of vortices
$r\lp t \rp$ as a function of re-scaled time $t$ {\it averaged}
over 9 experimental realisations (right).}
\end{center}
\end{figure}

In figure \ref{rdat}, we see the average separation between  vortices
$r(t)$ as a function of re-scaled time $t$ (left) and its logarithmic
derivative (right) for an specific experimental realisation. In figure
\ref{rfp} (left), we again see the corresponding periodogram of the
logarithmic derivative shown in figure \ref{rdat} with a peak at a
frequency of $\approx 4$. The height of the peak is again only approximately
twice the height of the second largest. Nevertheless,  performing the
average of  the Lomb periodograms for 9 experimental
realisations gives us the picture shown in figure \ref{rfp} (right). The
peak has again moved up slightly but is still clearly visible, which again
means that the position of a peak at a frequency of $\approx 4$ for the single
realisation was not accidental.

\subsubsection{Mean radius of vortices}

\begin{figure}
\begin{center}
\epsfig{file=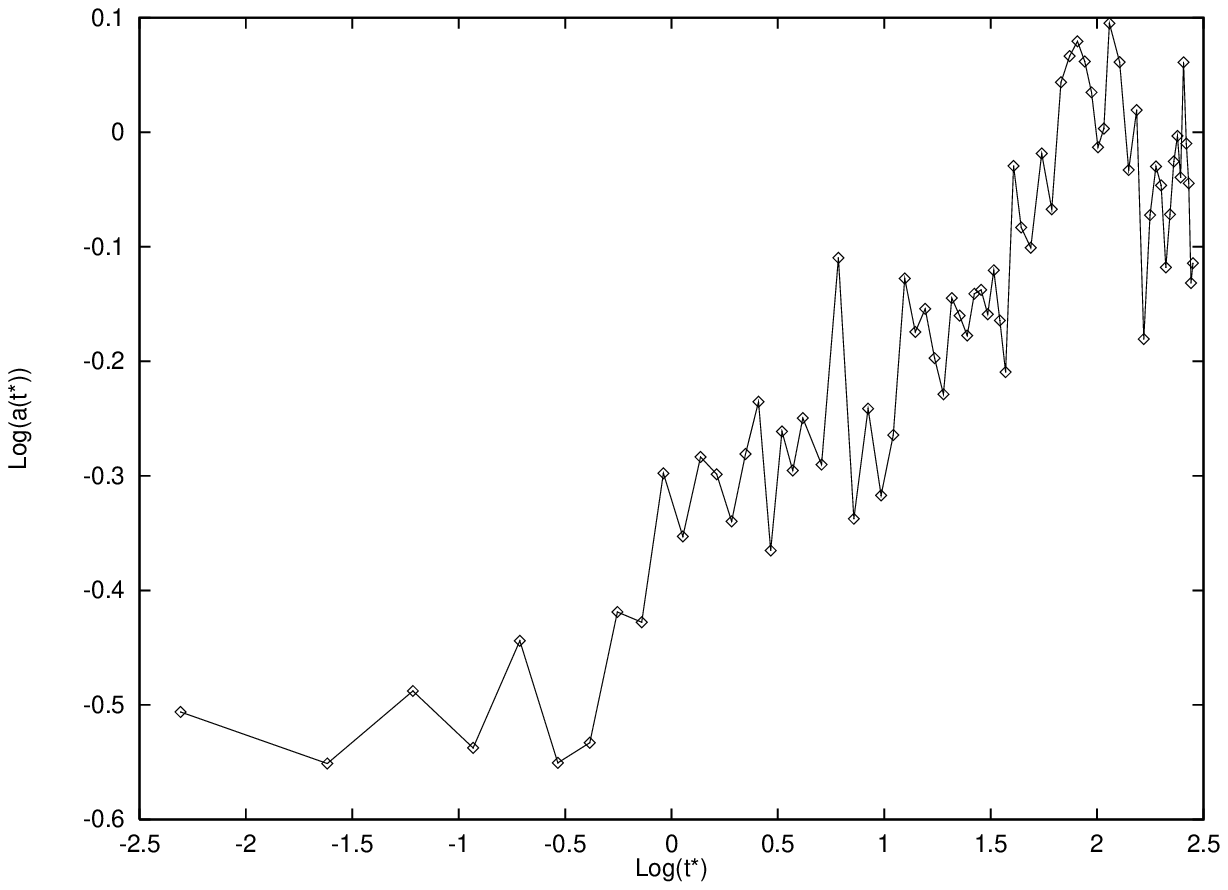,width=7.5cm}
\epsfig{file=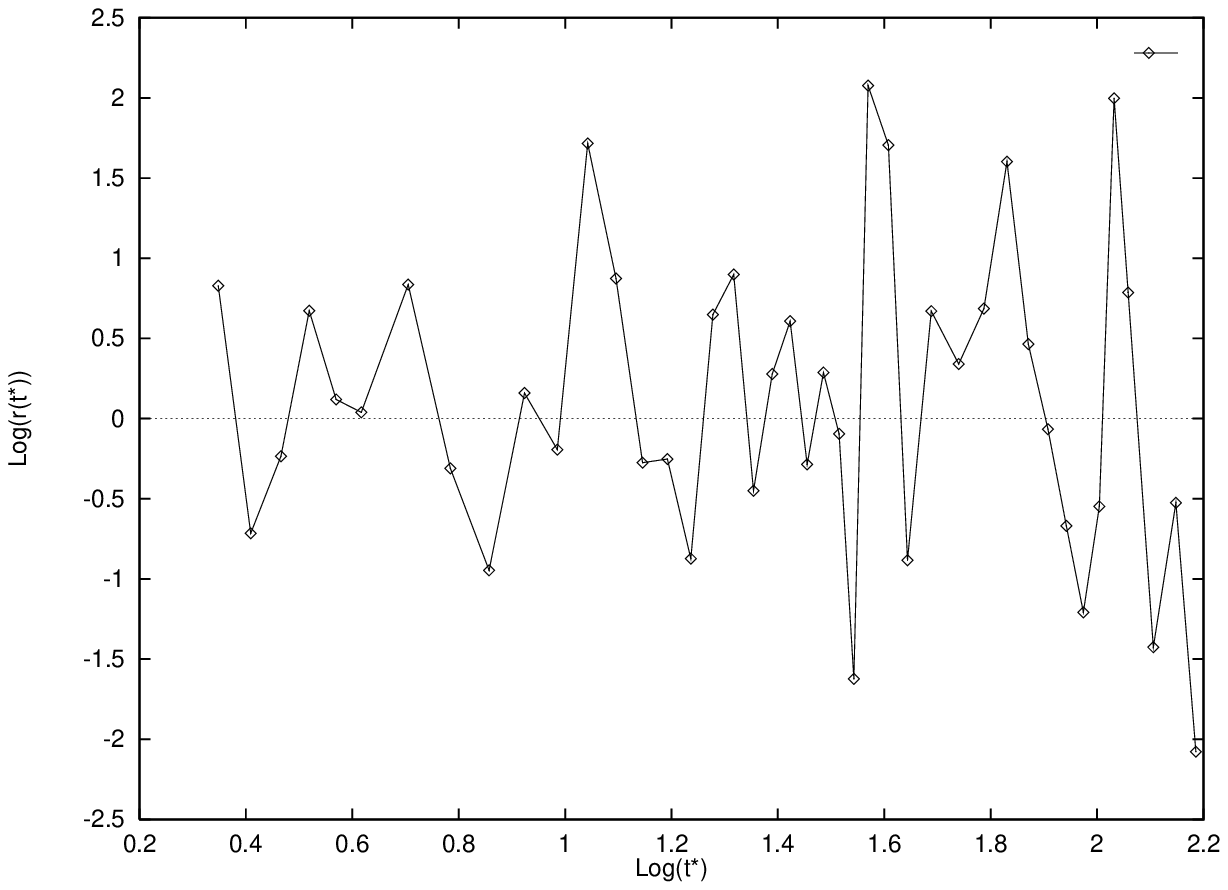,width=7.5cm}
\caption{\protect\label{adat} Logarithm of the mean radius of the vortices
$\log\lp a\lp t\rp \rp$ as a function of the logarithm of re-scaled time
$\log\lp t\rp$ (left). The corresponding logarithmic derivative as a
function of $\log\lp t\rp$.}

\vspace{1cm}

\epsfig{file=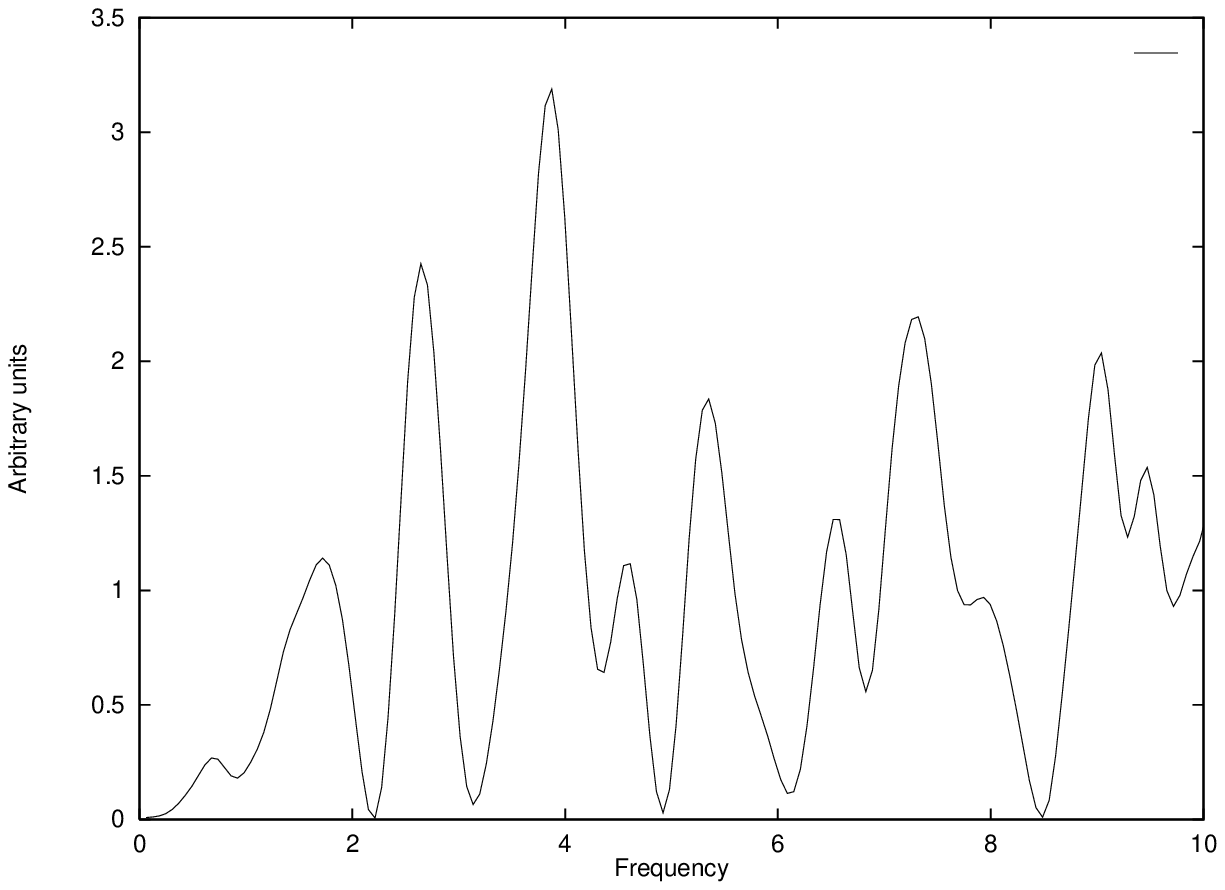,width=7.5cm}
\epsfig{file=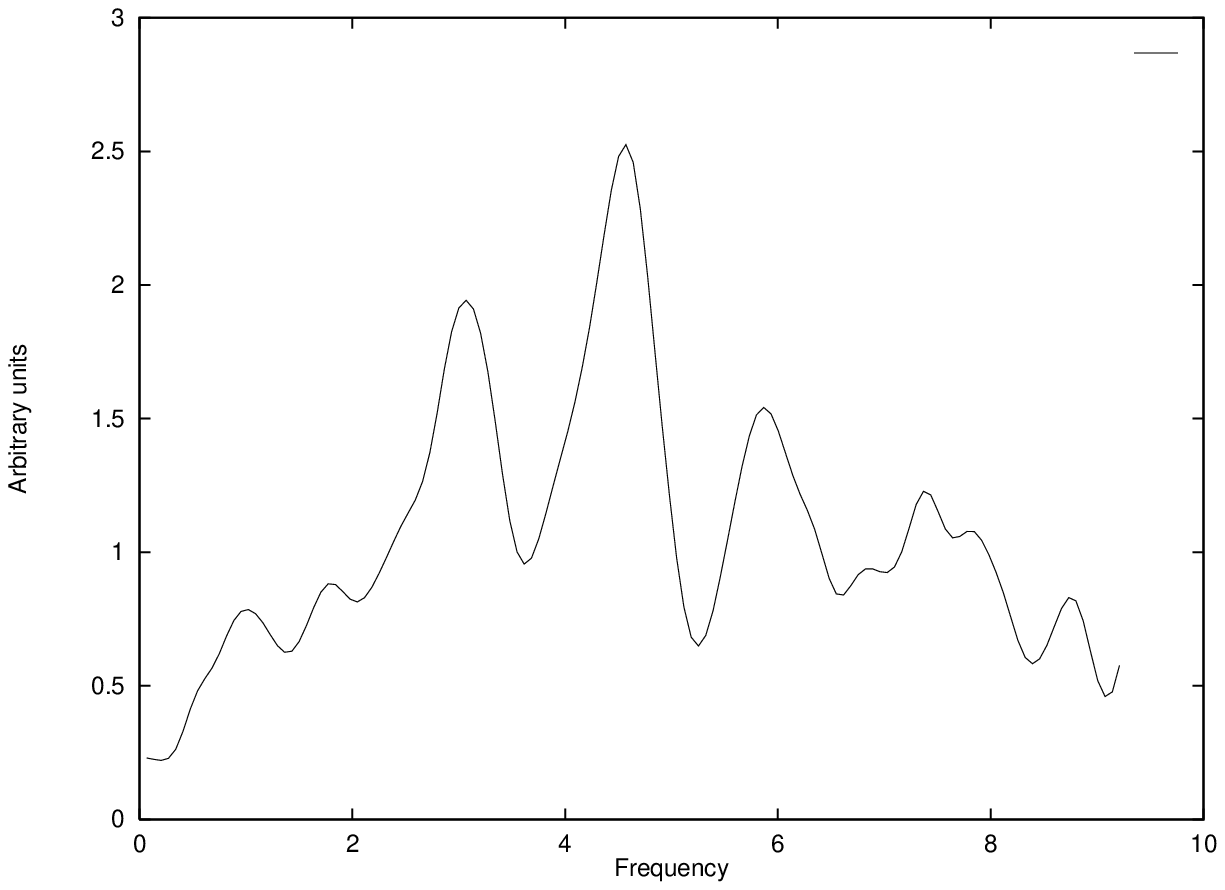,width=7.5cm}
\caption{\protect\label{afp} Lomb periodogram of the logarithmic
derivative shown in figure \protect\ref{adat} (left). Lomb periodogram
of the logarithmic derivative of the mean radius of the vortices
$a\lp t \rp$
as a function of re-scaled time $t$ {\it averaged} over 7 experimental
realisations (right).}
\end{center}
\end{figure}

In figure \ref{adat}, we see the mean radius of the vortices $a(t)$ as a
function of
re-scaled time $t$ (left) and its logarithmic derivative (right) for an
specific experimental realisation. In figure \ref{afp} (left), we see the
corresponding periodogram of the logarithmic derivative shown in figure
\ref{adat} with a peak at a (log-) frequency of $\approx 4$. The height
of the peak is again only approximately twice the height of the second
largest. Performing the average of the Lomb periodograms for 7
experimental realisations available here gives us the picture shown in
figure \ref{afp} (right). The peak is rather broad but is still clearly
visible, which again signifies that the position of a peak at a frequency of
$\approx 4$ for the single realisation was not accidental.

\section{Conclusion}

We have suggested that the concept of {\it discrete} time scale invariance
of the underlying physics
is more appropriate for the description of freely decaying 2-d turbulence
than the usual assumption of a continuous time scale invariance.

In this goal, a numerical analysis of synthetic data have demonstrated a
new canonical averaging scheme using the Lomb periodogram as its main
ingredient. We have furthermore shown that seemingly innocent data
manipulations completely destroy the log-periodic structures in these
synthetic data.

Using this new canonical averaging scheme on experimental data obtained
from freely decaying 2-d turbulence,
we have shown that the log-periodic signatures slightly visible in the
individual experimental realisations carry statistical significance and
cannot be disregarded as noise when averaged over the nine available
experimental realizations. It is quite reassuring that the analysis
of the three different experimentally independent measurements all agree on
a log-frequency of $\approx 4-5$ for the identified log-periodic oscillations.
This correspond to a prefered scale factor $\lambda \equiv e^{1/f} \approx
1.2-1.3$

Our study of this experiment establishes for the first time a systematic
procedure
for exploring the existence of log-periodic oscillations in turbulence
data. In this spirit,
it would be very interesting to analyse
previous partial indications of log-periodicity in turbulent data,
that can be found in fig. 5.1 ~p.58 and fig. 8.6~ p.128 of Ref.
\cite{Frisch}, fig.3.16~ p. 76 of Ref.\cite{Arneodo}, fig.1b of
Ref.\cite{Tcheou} and
fig. 2b of Ref. \cite{Castaing}.

For a theoretical view point, we have presented a theoretical framework
which makes plausible the presence of log-periodic
oscillations in the observables. Physically, we interpret the
observed
log-periodicity as resulting from a discrete cascade of intermittent
vortex coalescence processes. In this sense, 2-d freely decaying
turbulence would not be much
different from other systems in which log-periodicity has been documented
to result from self-similar intermittent instabilities \cite{revue}.

From the point of view of the Navier-Stokes equations, it may seem remarkable
or even stretching credibility that log-periodicity emerges from
such continuous PDE. A possible analogy is offered by
the recent discovery that the continuous nonlinear Einstein PDE's
of general relativity
in the presence of a scalar field self-interacting through gravitation may
generate
a log-periodic spectrum of black hole masses with develop according to a
log-periodic
self-similar time dynamics \cite{Blackhole}. The mechanism
might result from the existence of a limit cycle in the renormalisation
group description of
a field close to the negative density limit (in turbulence, could this be
obtained from
a negative effective viscosity?). Another possible route \cite{ET7}
is that scale invariant equations
that present an instability at finite wave-vector $k$ decreasing with the field
amplitude may generate naturally a self-similar discrete spectrum of
internal scales.

\vskip 0.5cm
We acknowledge useful comments from A. Arn\'eodo, B. Dubrulle, U. Frisch
and P. Tabeling.

\end{document}